\documentclass[aps,prl,twocolumn,showpacs,superscriptaddress,nobibnotes,nofootinbib,showkeys,10pt]{revtex4-2} 
\usepackage[dvips]{graphicx}
\usepackage{epsfig}
\usepackage{bm}   
\usepackage{dcolumn}
\usepackage{rotate}
\usepackage[table]{xcolor}
\usepackage{footmisc,booktabs,amssymb,longtable}
\usepackage{amsmath,amsfonts}

\usepackage{multirow}
\definecolor{Gray}{gray}{0.85}
\definecolor{LightCyan}{rgb}{0.88,1,1}

\usepackage[scaled=0.92]{helvet}
\usepackage[T1]{fontenc}
\usepackage{textcomp}
\usepackage{braket}
\usepackage{textgreek}
\usepackage{rotate}
\usepackage{tabularx}
\usepackage[referable]{threeparttablex}

\definecolor{capri}{rgb}{0.0, 0.75, 1.0}
\definecolor{cornflowerblue}{rgb}{0.39, 0.58, 0.93}
\definecolor{spirodiscoball}{rgb}{0.06, 0.75, 0.99}
\definecolor{pear}{rgb}{0.82, 0.89, 0.19}

\begin{document}

\title{Global trends of the electric dipole polarizability from shell-model calculations}

\author{Jos\'e Nicol\'as Orce}
\thanks{Corresponding author: jnorce@uwc.ac.za}
\email{coulex@gmail.com} \homepage{http://nuclear.uwc.ac.za}
\affiliation{Department of Physics \& Astronomy, University of the Western Cape, P/B X17, Bellville 7535, South Africa}
\affiliation{National Institute for Theoretical and Computational Sciences (NITheCS), South Africa}

\author{Cebo Ngwetsheni}
\affiliation{Department of Physics \& Astronomy, University of the Western Cape, P/B X17, Bellville 7535, South Africa}

\author{B. Alex Brown}
\affiliation{Department of Physics and Astronomy, and the Facility for Rare Isotope Beams, Michigan State University,
East Lansing, MI 48824-1321, USA}

\date{\today}

\begin{abstract}

Shell-model calculations of the  electric dipole ($E1$) polarizability have been performed for the
ground state  of selected {\it p-} and {\it sd-}shell nuclei, substantially advancing previous knowledge.
Our results are slightly larger compared with the somewhat more scattered photo-absorption cross-section data, albeit
agreeing with \emph{ab initio} calculations at shell closures and presenting a smooth
trend that follows the leptodermus approximation provided by the finite-range droplet model ({\small FRDM}).
The total $E1$ strengths also show an increasing trend proportional to the mass number which follows from the
classical oscillator strength ({\small TRK}) sum rule for the $E1$ operator.
The enhancement of the energy-weighted sum over $E1$ excitations with respect to the {\small TRK} sum rule arises from the use of experimental single-particle energies and the
residual particle-hole interaction.

\end{abstract}



\maketitle



\section{Motivation}

The ability for a nucleus to be polarized is driven by the dynamics of the isovector giant dipole resonance ({\small GDR})~\cite{baldwin1947photo}, which can
be characterized macroscopically as the collective motion of inter-penetrating proton and neutron fluids out of phase~\cite{migdal1945quadrupole,goldhaber1948nuclear,steinwedel1950nuclear},
and microscopically, through the shell-model ({\small SM}) interpretation of a system of independent nucleons plus configuration mixing
or the superposition of one particle - one hole (1p-1h) excitations~\cite{levinger1954independent,wilkinson1956nuclear,balashov1962relation,danos1965photonuclear}.
This collective motion can be related to the nuclear symmetry energy
$a_{_{sym}}(A)(\rho_{_N}-\rho_{_Z})^2/\rho_{_A}$ --- as defined in the Bethe--Weizs\"acker semi-empirical mass formula~\cite{weizsacker1935theorie,bethe1936nuclear} --- acting as a restoring force~\cite{berman1975measurements}, where $a_{_{sym}}(A)$ is the symmetry energy coefficient, $A$ the atomic mass number, $A=N+Z$, and $\rho_{_N}$, $\rho_{_Z}$ and $\rho_{_A}$ the neutron, proton and total fluid densities, respectively.

Within the hydrodynamic model,  the dipole polarizability $\alpha_{_{E1}}$
is directly proportional to $A^{5/3}$ and inversely proportional to $a_{_{sym}}(A)$~\cite{migdal1945quadrupole,levinger1957migdal},
\begin{equation}
 \alpha_{_{E1}} = \frac{P}{E}= \frac{e}{E}\int_V \rho^{\prime}z^2 dV = \frac{e^2R^2A}{40~a_{_{sym}}(A)},
 \label{eq:alphaasym}
\end{equation}
where $E$ is the magnitude of an electric field along the positive $z$ axis, $P$ the dipole moment with density $e\rho^{\prime}z^2$, $\rho^{\prime}=\frac{eE\rho_{_A}}{8a_{_{sym}}(A)}$, and $R$ the radius of a nucleus with a well-defined surface, $R=1.2~A^{1/3}$ fm.

Complementary, using non-degenerate perturbation theory, $\alpha_{_{E1}}$ is defined in terms of the energy-shift of the nuclear levels arising from the
quadratic Stark effect~\cite{flambaum2021nuclear}, i.e. $\Delta E = -\frac{1}{2} \alpha E^2$, and can be determined for
ground states with an arbitrary initial angular momentum $J_i$ using the inverse energy-weighted sum rule~\cite{deBoer1968},
\begin{eqnarray}
\alpha_{_{E1}}=\frac{2e^2}{2J_i+1}\sum\limits_{n} \frac{|\langle i\parallel\hat{E1}\parallel
n\rangle|^2}{E_n - E_i}=\frac{9\hbar c}{8\pi^3}\sigma_{_{-2}},
\label{eq:polar}
\end{eqnarray}
where $2J_i+1$ is the normalization constant arising from the Wigner-Eckart theorem~\cite{rose1957elementary,messiah1961quantum}, and
the sum extends over all $\lvert n\rangle$ intermediate states
connecting the initial ground state $\lvert i\rangle$ with electric dipole or $E1$ transitions.
The $\sigma_{_{-2}}$ value is the
$(-2)$ moment of the total 
photo-absorption cross section, $\sigma_{_{total}}(E_{\gamma})$,  defined as~\cite{levinger1960,migdal1965theory},
\begin{eqnarray}
\sigma_{_{-2}}=\int_{0}^{E_{\gamma}^{max}}
\frac{\sigma_{_{total}}(E_{{\gamma}})}{E_{{\gamma}}^{^2}}dE_{{\gamma}},
\label{eq:sigma-2int2}
\end{eqnarray}
which is generally integrated between particle threshold  and the available upper limit for monochromatic photons,
$E_{\gamma}^{max}\approx20-50$ MeV~\cite{dietrich1988atlas}.
An upper
limit of $E_{\gamma}^{max}\approx50$ MeV approximates the $\sigma_{_{-2}}$ asymptotic value for light and medium-mass
nuclei~\cite{ahrens1976experimental}. Magnetic polarizability contributions~\cite{knupfer1985scaling} are not considered here, but may be relevant for $^6$Li and $^7$Li~\cite{knupfer1981effect}.

The bulk of knowledge on how atomic nuclei polarize arises from photo-absorption cross-section data~\cite{dietrich1988atlas,plujko2018giant,kawano2020iaea}, where most of the absorption (and emission) of  photons is provided by the {\small GDR}~\cite{ishkhanov2021giant}.
Data predominantly  involve photo-neutron cross sections --- although photo-proton contributions are dominant for some light and $N=Z$ self-conjugate nuclei~\cite{orce2022competition} ---
and mainly concern the ground states of nuclei.
To a much lesser extend,  $\alpha_{_{E1}}$
has  been determined from several experiments using radioactive ion beams~\cite{rossi2013measurement}, inelastic proton scattering~\cite{tamii2011complete,roca2015neutron,hashimoto2015dipole,roca2018nuclear,bassauer2020evolution} and virtual photons~\cite{orce2020polarizability}.

\begin{figure*}[!ht]
\begin{center}
\includegraphics[width=8.2cm,height=6.8cm,angle=-0]{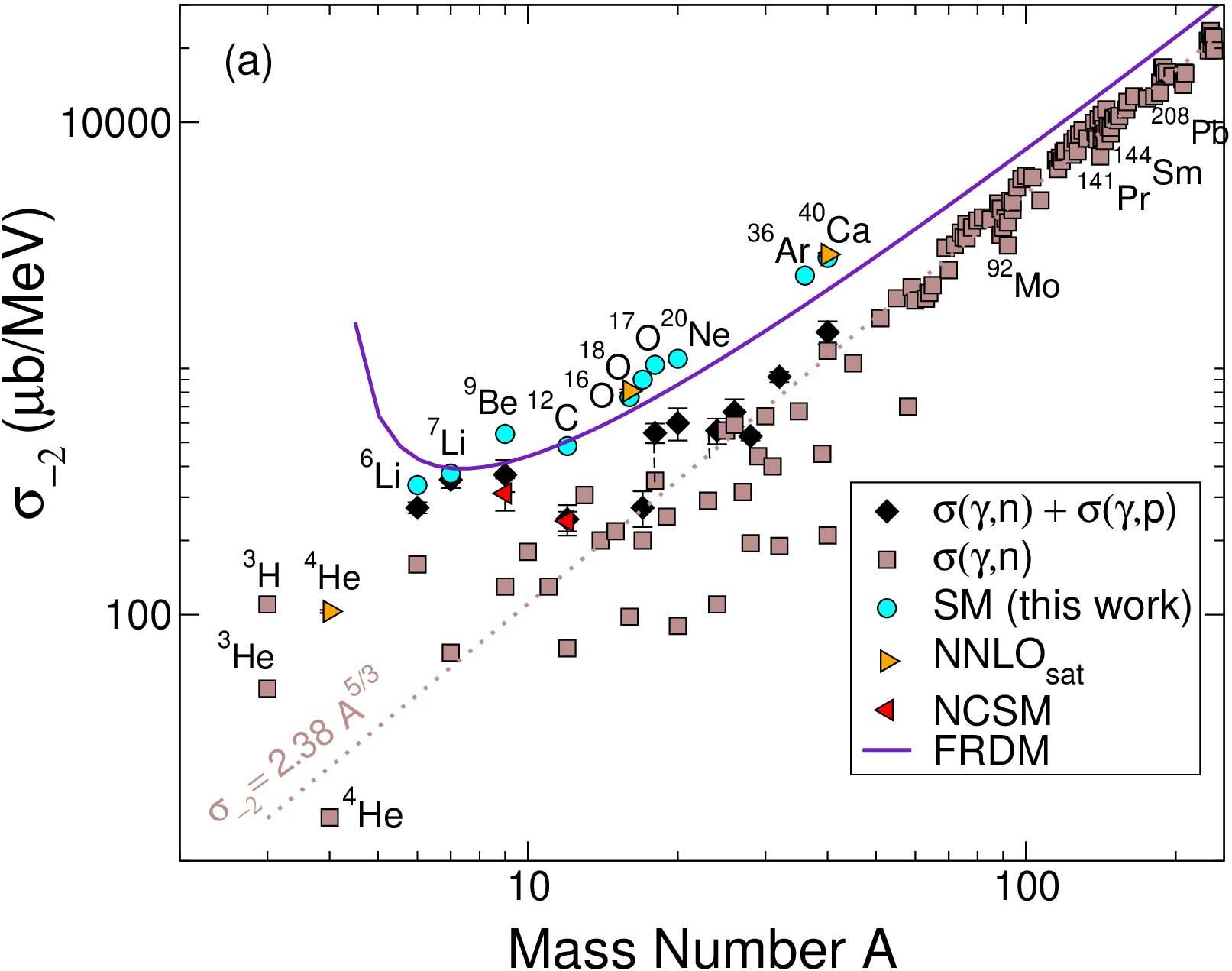}
\hspace{0.5cm}
\includegraphics[width=8.0cm,height=6.6cm,angle=-0]{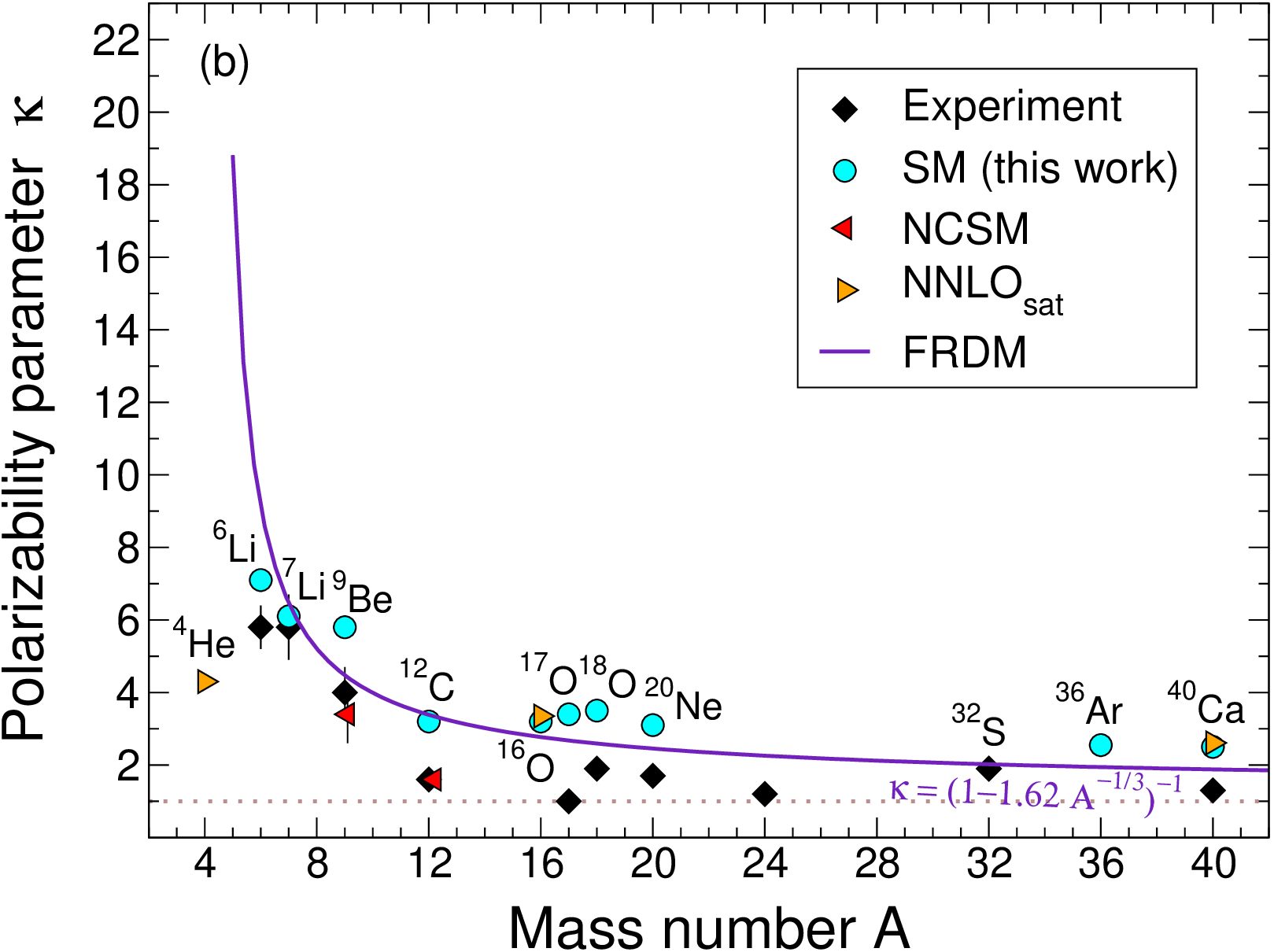}
\caption{Experimental and calculated $\sigma_{_{-2}}$ (left) and $\kappa$ (right) values for the ground states of selected {\it p} and {\it sd-}shell nuclei.  Data points (square and diamonds) arise from available photo-absorption cross-sections~\cite{dietrich1988atlas,nathans1953excitation,denisov1973photonuclear,berman1965photo,denisov1973photonuclear,bazhanov1965li6,kulchitskii1963energy,bramblett1973photoneutron,knupfer1981effect,fuller1985photonuclear,zubanov1992,jury1980,woodworth1979,allen1981,gorbunov1962,varlamov1979effect,anderson1969}. The theoretical results are taken from Refs.~\cite{miorelli2016electric} (coupled-cluster with {\small NNLO$_\mathrm{sat}$})
and \cite{orce2012reorientation,raju2018reorientation} (NCSM). The {\small WBP} and {\small FSU} SM calculations (circles) are from the current work. For comparison, the hydrodynamic-model prediction for $\kappa=1$ in Eq.~\ref{eq:2p4} is shown by dotted lines together with the leptodermous trend  (Eq.~\ref{eq:asym}) given by the {\small FRDM} symmetry energy coefficients, $S_v= 30.8$ MeV, $S_s/S_v = 1.62$ (solid lines)~\cite{moller1995atomic}.}
\label{fig:expsm}
\end{center}
\end{figure*}

Other phenomena that can contribute to $\sigma_{_{-2}}$ values are the pygmy dipole resonance ({\small  PDR})~\cite{gibelin2008decay,paar2007exotic,von2016comment,arsenyev2016effects,cook1957photodisintegration} --- an out-of-phase oscillation of the valence nucleons against the core ---
and low-energy nuclear resonances arising from cluster formation~\cite{eramzhyan1986giant,nakayama2001dipole,burda2010resonance}.
The {\small PDR} has been identified in light nuclei with neutron excesses~\cite{cook1957photodisintegration,zubanov1992,aumann1999giant,leistenschneider2001photoneutron,paar2007exotic,gibelin2008decay} --- being $^{13}$C where the ``pygmy resonance” was originally termed~\cite{cook1957photodisintegration} --- but has never been observed along the $N=Z$ line of stability.
Estimates suggest a contribution of about 5-10\% to $\sigma_{_{-2}}$ values, similar to those found in stable neutron-rich nuclei~\cite{von2016comment}.
A larger contribution is expected as $N$ increases and approaches the neutron drip line~\cite{terasaki2006self}.
An additional contribution could potentially arise from the low-energy enhancement ({\small LEE}) of the photon-strength function~\cite{burger2012nuclear,larsen2007nuclear,larsen2019novel,midtbo2021new,zilges2022photonuclear,ngwetsheni2019continuing,ngwetsheni2019combined,ngwetsheni2019how},
although $^{43-45}$Sc are the lightest nuclei
where it has been identified~\cite{burger2012nuclear,larsen2007nuclear}.


Photo-neutron cross-section data for $A\gtrapprox50$ nuclei --- where neutron emission is
generally the predominant decay mode --- show a smooth trend of  $\sigma_{_{-2}}$ values in the left panel of Fig.~\ref{fig:expsm}
(dotted line),  following the empirical power-law formula~\cite{orce2015new,orce2016reply},
\begin{eqnarray}
\sigma_{_{-2}}(A) = 2.38\kappa ~ A^{5/3}\,\mbox{$\mu$b/MeV},
\label{eq:2p4}
\end{eqnarray}
where $\kappa$ accounts for deviations from the actual {\small GDR} effects to
that predicted by the hydrodynamic model~\cite{migdal1945quadrupole,levinger1957migdal}.
Here, all quoted  $\sigma_{_{-2}}$ and $\kappa$ values are
related to Eq.~\ref{eq:2p4}.
Deviations from the smooth trend ($\kappa=1$) arise for loosely-bound light nuclei ($\kappa>1$)~\cite{orce2015new}
and semi-magic nuclei ($\kappa<1$)~\cite{ngwetsheni2019continuing,ngwetsheni2019combined,ngwetsheni2019how}, where the extra stability of shell closures may hinder polarization.

In this work, we investigate $\alpha_{_{E1}}$ for ground states of nuclei within the {\it p} and {\it sd} shells by performing novel $1\hbar\omega$ {\small SM} calculations --- following Eq.~\ref{eq:polar} --- with the {\small WBP} and {\small FSU} Hamiltonians.
We further explore deviations from the smooth trend ($\kappa=1$) predicted by the hydrodynamic model in Eq.~\ref{eq:2p4}, and compare our results with available photo-absorption cross-section data~\cite{exfor,ENDF}, sums of $E1$ strengths and the classical oscillator strength sum rule for the $\hat{E1}$ operator.

\begin{table*}[!ht]
\begin{center}
\caption{Experimental and calculated $\sigma_{_{-2}}$ (columns 3 and 4) and $\kappa$ (columns 7 and 8) values of ground states in selected $p$-$sd$ shell nuclei.
Experimental data arise from available photo-absorption cross-sections. Previous calculations (column 9) are listed for comparison.
All quoted  $\sigma_{_{-2}}$  and $\kappa$ values presented in this work are related to Eq. 4.\label{tab:tab1}}
\begingroup
\setlength\textwidth\textheight
\setlength{\tabcolsep}{4pt}
{ \footnotesize
\begin{tabular}{|ccccccccc|}
\hline \hline
Nucleus & $J^{\pi}$  &  $\sigma_{_{-2}}^{exp}$ &  $\sigma_{_{-2}}^{SM}$ &  $E_{_{max}}^{^{SM}}$  & $\sum\limits_{n} {B(E1_n)^{SM}}$  & $\kappa^{exp}$   & $\kappa^{SM}$  &  $\kappa^{previous}$ \\
        &            &  $\mu$b/MeV             &   $\mu$b/MeV           &        MeV             &         $e^2fm^2$        &                  &                &           \\
\hline
\hline
${^6}$Li    & 1$^+_{_1}$    &   272(14)~\cite{knupfer1981effect,denisov1973photonuclear,berman1965photo,bazhanov1965li6}  &  336   &  34.0  &  1.7   & 5.8(6)         &  7.1         &   --    \\
\hline
${^7}$Li    & 3/2$_{_1}^{-}$   &  353(26)~\cite{knupfer1981effect,kulchitskii1963energy,bramblett1973photoneutron}   &  374  & 47.0  & 1.9   & $5.8(9)$  &  6.1 &   --       \\
\hline
${^9}$Be    & 3/2$_{_1}^{-}$   & 370(55)~\cite{nathans1953excitation}    &  542  &  58.3  & 2.5  & 4.0(8)   &  5.8  & 3.4(8)~\cite{orce2012reorientation} \\
\hline
${^{12}}$C  & 0$_{_1}^{+}$     & 244(35)~\cite{fuller1985photonuclear}    & 484 & 65.1  &  2.9  & 1.6(2)   & 3.2  & 1.6(2)~\cite{raju2018reorientation} \\
\hline
${^{16}}$O  & 0$_{_1}^{+}$    & 616(90)~\cite{fuller1985photonuclear}   &  765  & 25.9 & 4.5  &  2.5(4)  & 3.2  &  3.4(1)~\cite{miorelli2016electric}     \\
\hline
${^{17}}$O  & 5/2$_{_1}^{+}$    & 272(45)~\cite{zubanov1992,jury1980}   & 901  & 35.2 &  4.7  & 1.2(2)  & 3.4  &  --     \\
\hline
${^{18}}$O  & 0$_{_1}^{+}$     & 547(50)~\cite{woodworth1979}   &  1035 & 44.3 & 5.3  & 1.9(3)        & 3.5  &   --     \\
\hline
${^{20}}$Ne & 0$_{_1}^{+}$    & 600(90)~\cite{allen1981,gorbunov1962}   & 1095  & 47.0 & 6.3 &  1.7(3)   & 3.1   &   --    \\
\hline
${^{24}}$Mg & 0$_{_1}^{+}$     & 559(66)~\cite{varlamov1979effect,anderson1969}   &  1132 & 42.2 & 5.9  & 1.2(2) &  2.4   &   --    \\
\hline
${^{36}}$Ar & 0$_{_1}^{+}$     &   --      & 2384 & 31.8  &  11.6  &  --                        & 2.6   &   --    \\
\hline
${^{40}}$Ca & 0$_{_1}^{+}$     &  1405(150)~\cite{goryachev1968cross,goryachev1968structure}   & 2813 & 24.5  &  13.8  &  1.3(2)                        & 2.5   &   2.6(1)~\cite{miorelli2016electric}    \\
\hline \hline
\end{tabular}
}\endgroup
 \begin{tablenotes}
      \small
      \item Shell model calculations below $^{17}$O are from the {\small WBP} interaction whereas for $A\geq17$ we quote the values from the {\small FSU} interaction.
      Slightly smaller values are determined with the  {\small WBP} interaction in the middle and end of the {\it sd} shell.
    \end{tablenotes}
\end{center}
\end{table*}

\section{Shell-model Calculations}

Shell-model calculations of the $E1$ polarizability are demanding since they normally involve hundreds of $E1$ matrix elements and high-performance computing. Priorly, {\small SM} calculations of  $\kappa$ values for ground states have been carried out --- following Eqs.~\ref{eq:polar} and \ref{eq:2p4} --- in $^{9,10}$Be~\cite{orce2012reorientation} and $^{12}$C~\cite{raju2018reorientation} using the no-core shell-model ({\small NCSM}) with the {\small CD}-Bonn $2N$ and
 chiral effective field theory ({\small $\chi$EFT}) $2N + 3N$ forces~\cite{navratil2007local,roth2014evolved,entem2003accurate,entem2017high,entem2015peripheral,bogner2007similarity},
 $N_{max}=4$ basis sizes for natural and $N_{max}=5$ for  unnatural parity states. These {\it ab initio} calculations included $E1$ matrix elements from all
the transitions connecting about 30 1$^-$ states up to 30 MeV in $^{10}$Be and $^{12}$C, 
and $E1$ contributions from about 100 intermediate 1/2$^+$, 3/2$^+$, and 5/2$^+$ states in $^{9}$Be.
For the ground states of $^{9}$Be and $^{12}$C, values of $\kappa(g.s.)=3.4(8)$ and $\kappa(g.s.)=1.6(2)$ 
were predicted, respectively, in agreement with photo-absorption cross-section data~\cite{levinger1957migdal,nathans1953excitation,fuller1985photonuclear}. Additional {\small NCSM} calculations of
$\alpha_{_{E1}}$ values  have  been performed in $^3$H, $^3$He, and $^4$He by Stetcu and collaborators
using directly the Schr\"odinger equation and {\small $\chi$EFT} interactions~\cite{stetcu2009electric}.\\

In the present work,  $1\hbar\omega$ {\small SM} calculations of the ground-state $E1$ polarizability have been carried out using the {\small OXBASH} code~\cite{brown1988computer} with the Warburton and Brown ({\small WBP})~\cite{warburton1992effective} and Florida State University ({\small FSU})~\cite{lubna2019structure,lubna2020evolution,brown2022nuclear}
Hamiltonians within the {\it spsdpf} model space.
For nuclei near $^{16}$O,  the {\small FSU} Hamiltonian~\cite{lubna2019structure,lubna2020evolution,brown2022nuclear} is the same as the {\small WBP}
Hamiltonian from Ref.~\cite{warburton1992effective}. The single-hole energies are fixed
to the experimental separation energies between the $^{16}$O ground state
and states in $^{15}$O;
$-22.11$ and $-15.54$ MeV for $  0p_{1/2}  $ and $  0p_{3/2}  $, respectively.
The single-particle states are determined by
the separation energies between states in $^{17}$O
and the $^{16}$O ground state, as used in the {\small USDB} Hamiltonian \cite{brown2006new}, $-3.93$, $-3.21$ and 2.11 MeV for
$  0d_{5/2}  $, $  1s_{1/2}  $ and $  0d_{3/2}  $, respectively.
The energies of the pure  1p-1h  states range from  11.6
to 24.1 MeV.

The two-body matrix elements ({\small TBME})
involving both $  0p  $ and $  0d-1s  $ in the {\small WBP} Hamiltonian
for nuclei near $^{16}$O were
obtained from a realistic potential model that contains
a fixed one-pion exchange part, plus adjusted strengths
of central, spin-orbit and tensor
contributions from one-boson exchange (see Eq. 4 in \cite{warburton1992effective}).
There is no explicit  velocity dependence~\cite{levinger1950dipole}.
The parameters were obtained by fitting to the data
given in Table IV of \cite{warburton1992effective}.
The {\small FSU} Hamiltonian starts with the {\small WBP} Hamiltonian, and then
adjusts linear combinations of {\small TBME} of the type
$  \langle 0d-1s, 0f-1p, J, T \mid V \mid  0d-1s, 0f-1p, J, T \rangle  $
to fit the data shown in Fig. 6 of Ref.~\cite{lubna2019structure} ---
which include
particle-hole states originating from cross-shell excitations
that give rise to intruder states ---
and is built upon tuning the monopole terms across the shell gaps $N = 8$ and $N = 20$ to reproduce the experimental data.


\begin{figure}[!ht]
\begin{center}
\includegraphics[width=7.cm,height=5cm,angle=-0]{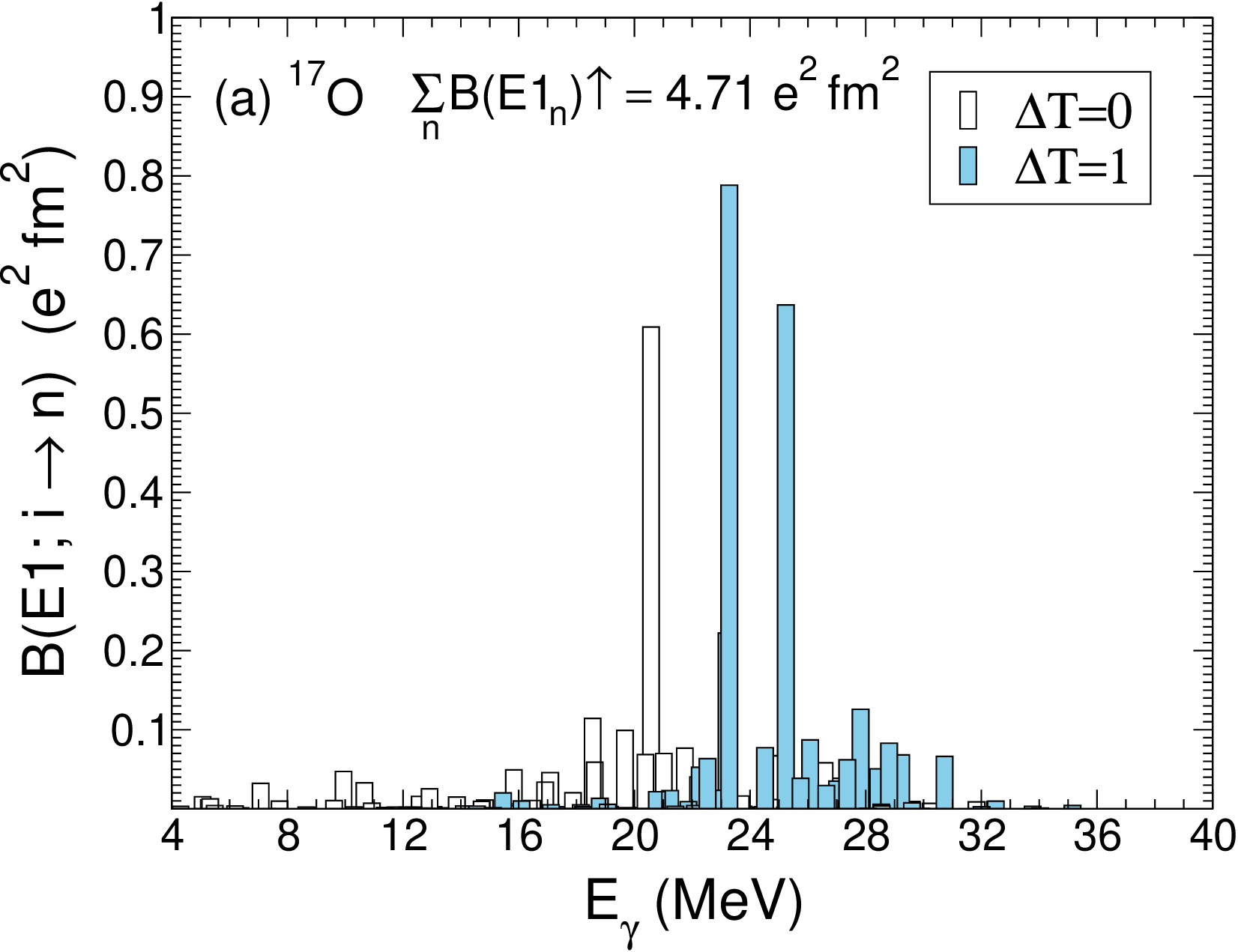} \\
\vspace{0.1cm}
\includegraphics[width=7.cm,height=5cm,angle=-0]{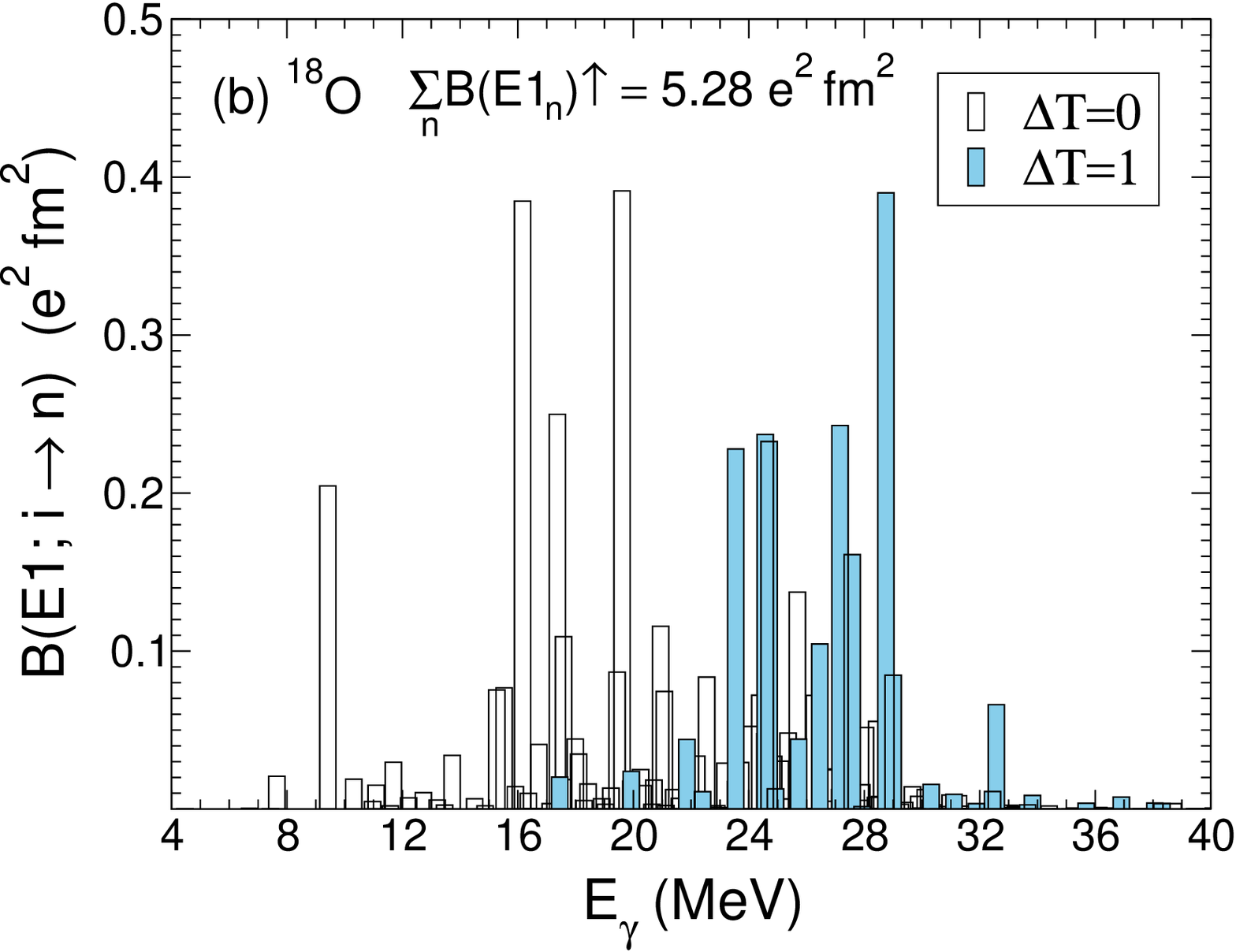}
\caption{Calculated $B(E1;0^+_{_1} \rightarrow 1^-_n)$ isovector distribution as a function of transition energies for $^{17}$O and $^{18}$O for both $\Delta T=0$ and $\Delta T=1$ transitions.~\label{fig:isospinsplitting}}
\end{center}
\end{figure}

\begin{figure*}[!ht]
\begin{center}
\includegraphics[width=6.cm,height=4.5cm,angle=-0]{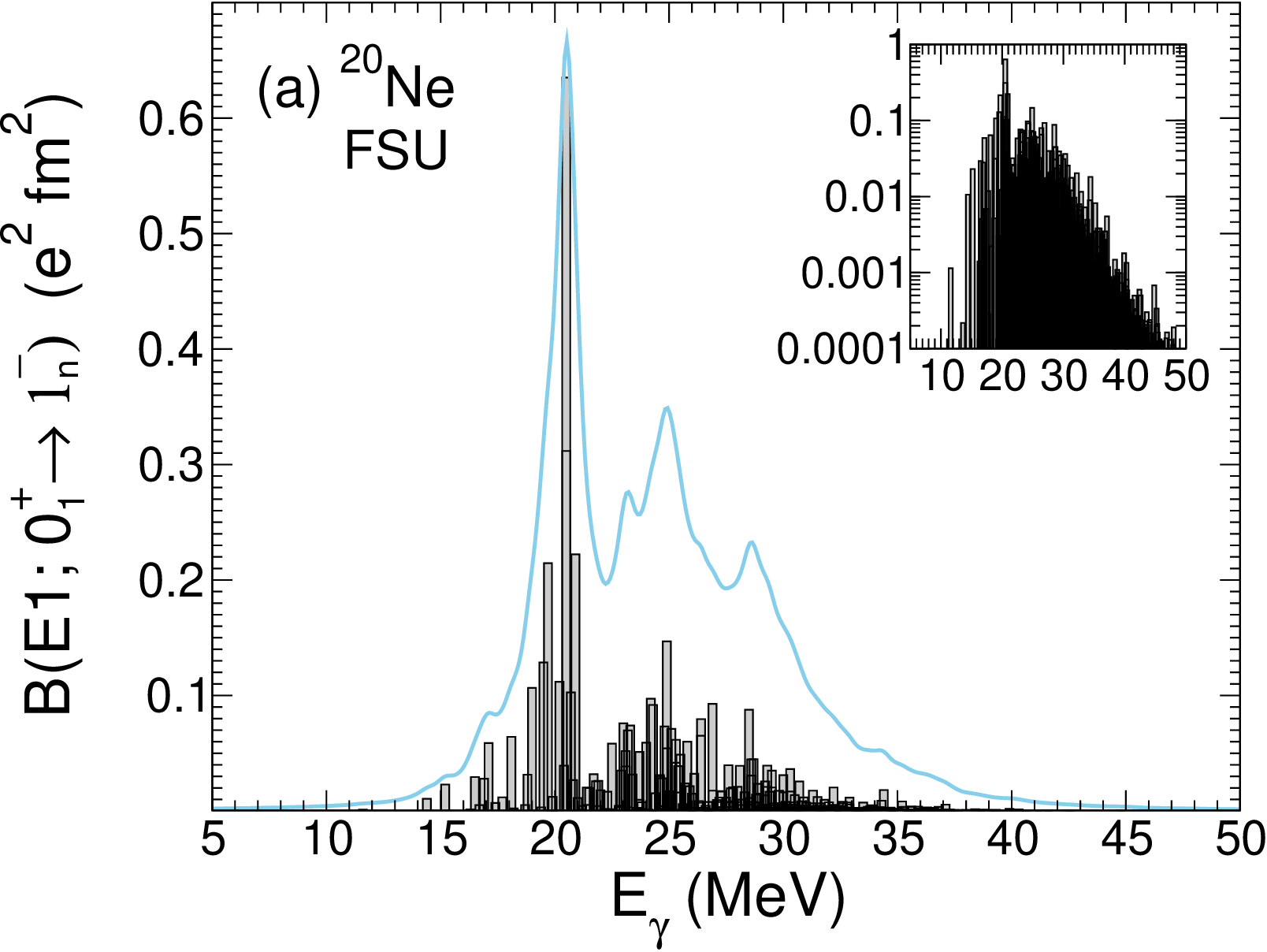}
\hspace{0.0cm}
\includegraphics[width=5.7cm,height=4.5cm,angle=-0]{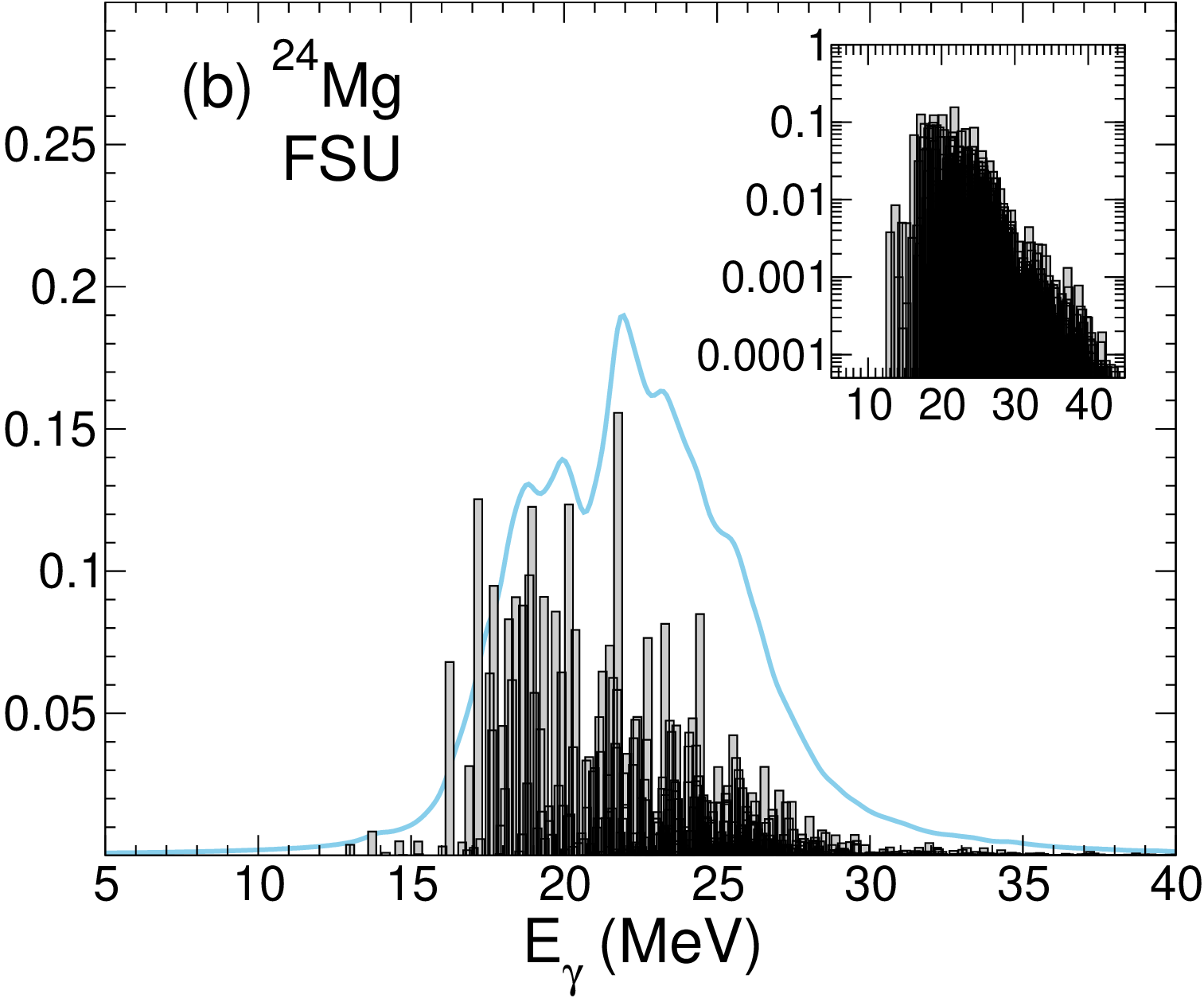}
\hspace{0.0cm}
\includegraphics[width=5.7cm,height=4.5cm,angle=-0]{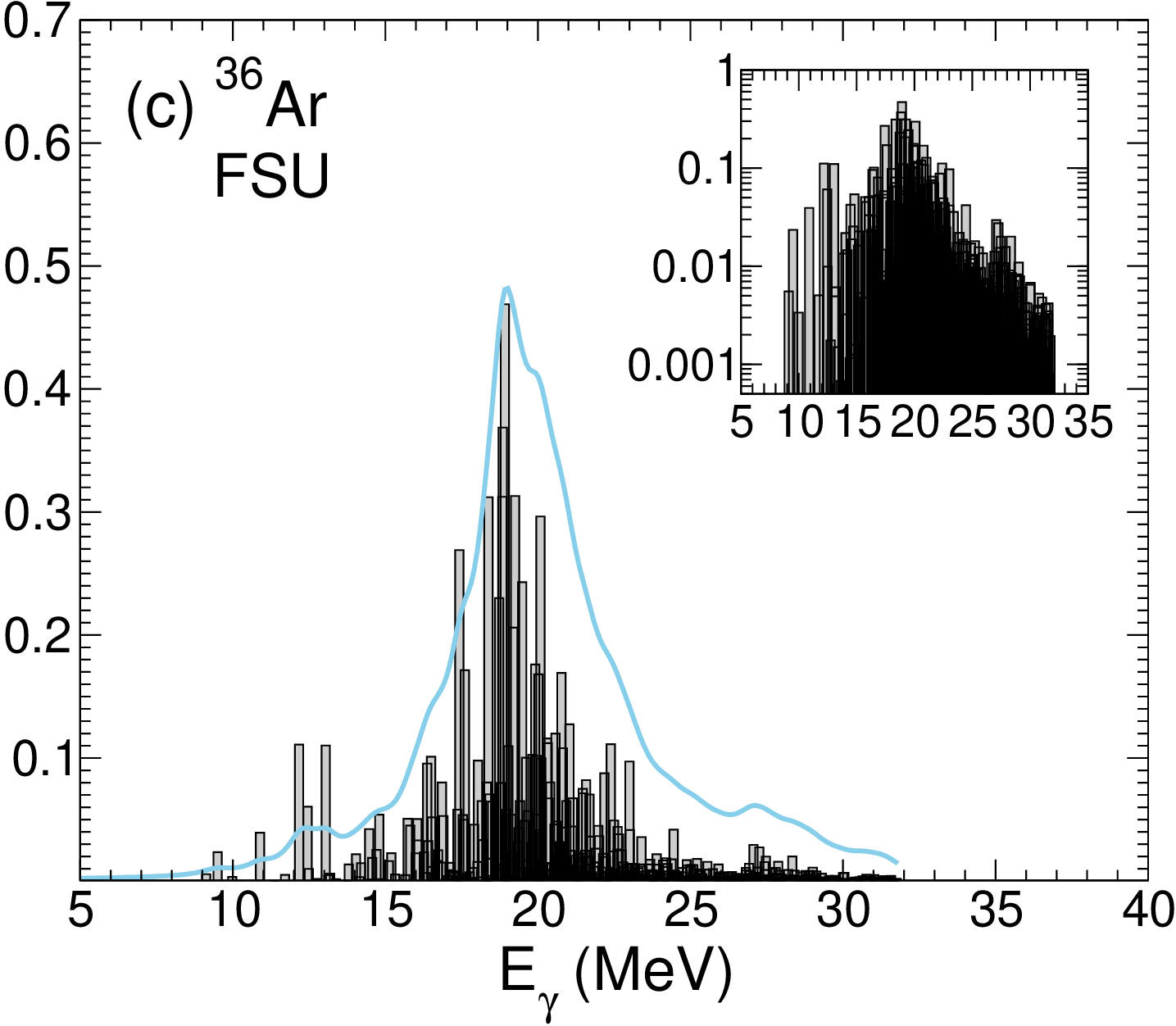}
\caption{Calculated $B(E1;0^+_{_1} \rightarrow 1^-_n)$ isovector distribution ($\Delta T =1$) as a function of transition energies for $^{20}$Ne, $^{24}$Mg and $^{36}$Ar. The light blue curves represent the strength distributions folded with a Lorentzian function with 1-MeV width, and illustrate the spreading of the strength.~\label{fig:20Ne24Mg36Ar}}
\end{center}
\end{figure*}

The general procedure in our {\small SM} calculations is calculating all $E1$ matrix elements following Eq.~\ref{eq:polar}.
For even-even nuclei, we calculate $\langle 0^+_{1} \parallel\hat{E1}\parallel 1^-_{n}\rangle$ $E1$ matrix elements connecting the ground state  with
up to 1000 intermediate $1^-_{n}$ isovector excitations with a maximum calculated energy  $E_{_{max}}^{^{SM}}$ listed in Table~\ref{tab:tab1}, which
include the {\small GDR} region. Similarly, we calculate all possible $E1$ matrix elements from
the various intermediate states for odd-even and odd-odd nuclei. {\small  SM} calculations for $^{28}$Si and $^{32}$S were not feasible because of limiting computing power and accuracy.

Isospin selection rules for $E1$ transitions~\cite{warburton1969role} have to be considered
according to the corresponding Wigner $3j$ symbols;   ergo,  isovector contributions for $N=Z$ self-conjugate nuclei ($T_{_z}=\frac{N-Z}{2}=0$) arise only from $\Delta T = 1$ transitions,
whereas both $\Delta T = 0,  1$ isovector transitions have to be considered otherwise. Such an isospin splitting  of $E1$ strengths
is shown in Fig.~\ref{fig:isospinsplitting} for $^{17}$O (top) and $^{18}$O (bottom),
where $B(E1_n)$ is the reduced transition probability connecting the ground state $|i\rangle$ with each final state $|n\rangle$~\cite{bohr1998nuclear},
\begin{equation}
 B(E1_n) = B(E1; i \rightarrow n) = \frac{1}{2J_i +1} ~|\langle n || \hat{E1} || i \rangle|^2.
\end{equation}
Isospin mixing at high-excitation energies is less than 5\% in the {\small GDR}
region~\cite{morinaga1955effects,barker1957effect,eramzhyan1986giant}.

Moreover, $E1$ effective charges are not required since all the $E1$ matrix elements are calculated in the full $1\hbar\omega$ model space, i.e.,
as shown in Fig.~\ref{fig:20Ne24Mg36Ar}, we do a full and fully converged $1\hbar\omega$ calculation of 1p-1h excitations that occur between major shells.
Dipole excitations far above the {\small GDR} region have a negligible effect.
Although $3\hbar\omega$ 1p-1h matrix elements are all zero in the harmonic oscillator ({\small H.O.}),
novel {\small SM} calculations by Sieja  in the neon isotopes show that possible admixtures with $1\hbar\omega$ 1p-1h $+$ 2p-2h transitions may suppress the $E1$ strength by about 15\%~\cite{Sieja2022,sieja2023brink}.

Removing spurious states is of particular relevance for $E1$ transitions in $N\approx Z$ nuclei because the motion of a particle involves the recoil of the rest of the nucleus, with the total center of mass remaining at rest~\cite{bohr1998nuclear}. Following the Gloeckner-Lawson method~\cite{gloeckner1974spurious},
the center-of-mass Hamiltonian incorporated in {\small OXBASH} conveniently pushes away 1p-1h spurious states involving $0s$ to $0p$, $0p$ to $0d1s$ and $0d1s$ to $0f1p$, and decouples them from intrinsic excitations~\cite{brown2005lecture}.
Table~\ref{tab:tab1} and Fig.~\ref{fig:expsm}
show the calculated $\sigma_{_{-2}}^{{SM}}$ and $\kappa^{{SM}}$ values for the ground states of selected $p$ and {\it sd-}shell nuclei.
For comparison, other theoretical results are also presented together with available photo-absorption cross-section measurements of $\sigma_{_{-2}}^{{exp}}$ and $\kappa^{{exp}}$ values.
It should be noted that previous {\it ab initio} calculations of $\alpha_{_{E1}}$  in $^{16}$O and $^{40}$Ca~\cite{miorelli2016electric} using integral transforms and the coupled-cluster method with the NNLO$_\mathrm{sat}$ interaction (right triangles in Fig.~\ref{fig:expsm}) agree with our results.


\section{Results and Discussion}

Pronounced deviations from  hydrodynamic-model estimates (dotted $\kappa=1$ lines in Fig.~\ref{fig:expsm}) are  calculated
for the ground states of $^{6,7}$Li and $^9$Be,  with slightly larger $\sigma_{_{-2}}$ and $\kappa$ values than those found experimentally.
Such deviations from the {\small GDR} effect are not surprising considering
the fragmentation of the {\small GDR} spectrum into different 1p-1h states~\cite{eramzhyan1986giant}, which include the possibility of
$\alpha$ cluster configurations~\cite{neudatchin1979clustering,he2014giant}  and the virtual breakup into the continuum~\cite{smilansky1972role,weller1985electromagnetic}.
The latter even led Smilansky, Weller and co-workers to suggest that the main contribution to the polarizability in $^7$Li may not actually be the {\small GDR},
but instead the virtual breakup into the $\alpha$-t continuum.
The rise of cluster structures in $^{6,7}$Li, $^9$Be as well as $^{17}$O can be pinned down to the extra loosely-bound or slightly unbound particle~\cite{barke1989investigation,barker1984decay}
--- whose wave function  extends far apart from the $\alpha$-cluster configurations, i.e. $\alpha+d$, $\alpha+t$, $2\alpha+n$, $4\alpha+n$, respectively --- as inferred from the dipole resonances observed at relatively low-excitation energies~\cite{zubanov1992,nakayama2001dipole,burda2010resonance}. Deviations from the hydrodynamic model in self-conjugate $N=Z$ nuclei
could also arise because of cluster formation~\cite{ikeda1968systematic,kanada1995clustering,von2006nuclear,freer2007clustered,ebran2012atomic,zhou2012new,he2014giant}
and/or the missing admixtures with $1\hbar\omega$ 1p-1h + 2p-2h transitions~\cite{Sieja2022,sieja2023brink}.
Despite slightly larger values are generally calculated compared to measurements, the right panel of
Fig.~\ref{fig:expsm} present similar theoretical and experimental trends of $\kappa$ values, and suggest that the bulk of these effects are implicitly incorporated in the phenomenological Hamiltonians.
The larger $\kappa^{SM}$  value in $^{17}$O is about 3 times larger than the experimental one and deserves further investigation as a
substantially larger  $\kappa=8.4(6)$ has also been experimentally determined for its first excitation~\cite{kuehner1982measurement}.

Interestingly, the overall smooth trend of $\sigma_{_{-2}}^{SM}$ and $\kappa^{SM}$ values shown in Fig.~\ref{fig:expsm} can be independently correlated with the leptodermous approximation ($A^{-1/3} \ll 1$) of the symmetry energy~\cite{tian2014effect,myers1969average,orce2015new},
\begin{equation}
 a_{_{sym}}(A)=S_v\left( 1-\frac{S_s}{S_v}~A^{-1/3} \right),
 \label{eq:asym}
\end{equation}
where $S_v$ is the volume symmetry-energy coefficient
and $\frac{S_s}{S_v}$ the surface-to-volume ratio for finite nuclei.


The set of $S_v$  and $S_s$ parameters that best reproduce the overall trends in Fig.~\ref{fig:expsm} is provided by the finite-range droplet model ({\small FRDM})~\cite{moller2016nuclear}, i.e.  the combination of the finite-range droplet macroscopic model and the folded-Yukawa single-particle microscopic model~\cite{moller1995atomic}.
The leptodermus trends for $\sigma_{_{-2}}$ and $\kappa$ are shown in  Fig.~\ref{fig:expsm} (solid lines) and determined by combining the {\small FRDM} parameters ($S_v= 30.8$ MeV and $S_s/S_v = 1.62$) with Eqs.~\ref{eq:alphaasym}, ~\ref{eq:polar}, and ~\ref{eq:2p4},
\begin{eqnarray}
\sigma_{_{-2}}(A) &=&  ~\frac{2.35 ~A^2}{A^{1/3}-1.62},  \\
\kappa(A) &\approxeq& ~\frac{1}{1-1.62A^{-1/3}},
\end{eqnarray}
which characterize the enhancement observed for light nuclei.
Additional sets of $S_v$ and $S_s/S_v$ parameters are discussed in Ref.~\cite{orce2020polarizability} and references therein, albeit presenting
larger deviations from the overall trends of {\small SM} calculations and data.\\

Further support of {\small SM} calculations may arise from  the calculated sum of
$E1$ strengths, $\sum\limits_{n}{B(E1_n)}$, shown in Tables~\ref{tab:tab1} and \ref{tab:BE1s}. The
$B(E1_n)$ values are calculated with {\small H.O.} radial wave functions.
The simplest approximation is $\hbar\omega = 41~ A^{-1/3}$ MeV for uncorrelated or independent-particle motion of the nucleons.
That is, when there is no particle-hole interaction, the 1$^{-}$, $T=1$ {\small GDR}
is split among all possible 1p-1h  states at an energy of $E_{_{GDR}} = 1 \hbar\omega$.
The total $E1$ strength then obeys the classical oscillator strength or {\small TRK}  sum rule~\cite{thomas1925zahl,ladenburg1923absorption,reiche1925zahl,kuhn1925gesamtstarke,bohr1998nuclear},
\begin{eqnarray}
{S}(E1)^{^{TRK}} &=&  E_{_{GDR}} ~\Sigma  B(E1_n) \\ \nonumber
&=& 14.8 ~\frac{NZ}{A} ~\mbox{MeV}~e^2~\mbox{fm}^2.
\label{eq:SE1class}
\end{eqnarray}
For  $N=Z$, ${S}(E1)^{^{TRK}}= 3.7 A$ MeV e$^{2}$ fm$^{2}$,
and $\Sigma  B(E1_n)^{^{TRK}} = 0.090 A^{4/3}$ e$^{2}$ fm$^{2}$ when  $  E_{_{GDR}} = 1 \hbar\omega$.
For our calculations we use the $\hbar\omega$ required to reproduce the $rms$ charge radii~\cite{bertsch1972practitioner}, which are listed in Table~\ref{tab:BE1s} ---
in closer agreement with Blomqvist and Molinari's $\hbar \omega  = 45 A^{-1/3} - 25 A^{-2/3}$ MeV~\cite{blomqvist1968collective} --- together with the
corresponding $\Sigma  B(E1_n)^{rms}$ for $N=Z$.
For example, for  $\hbar\omega$ = 13.26 MeV in $^{16}$O, there are five 1p-1h 1$^{-}$, $  T=1  $ states with $  \Sigma  B(E1_n)  $ = 4.47 e$^{2}$ fm$^{4}$.

\begin{table}[!ht]
\begin{center}
\caption{Classical  ${S}(E1)^{^{TRK}}$ (Eq.~\ref{eq:SE1class})
and $S(E1)^{^{SM}}$ (Eq.~\ref{eq:SE1wbp})  sum rules.
The {\small H.O.} energy $\hbar \omega$ in column 4  is fit to reproduce the $rms$ charge radii --- instead of using $\hbar\omega = 41~ A^{-1/3}$ MeV --- providing $\Sigma  B(E1_n)^{rms}$ for $N=Z$ nuclei without particle-hole interactions. The $f_{_{GDR}}$ ratio indicates the fraction of the {\small TRK} sum rule exhausted for each nucleus.~\label{tab:BE1s}}
\footnotesize
\begin{tabular}{|cccccc|}
\hline \hline
Nucleus   &  ${S}(E1)^{^{TRK}}$  &  ${S}(E1)^{^{SM}}$   &   $\hbar \omega$   &  $\Sigma  B(E1_n)^{rms}$  & $f_{_{GDR}}$     \\
          &    e$^2$fm$^2$MeV    &    e$^2$fm$^2$MeV     &    MeV             &       e$^2$fm$^2$     &     \\
\hline
\hline
$^6$Li    &     22.2      &  36.2       &  11.74  &  1.89   &  1.63                        \\
$^7$Li    &     25.4      &  42.5       &  13.71  &  1.85   &  1.67                        \\
$^9$Be    &     32.9      &  51.1       &  13.51  &  2.43   &  1.55                        \\
$^{12}$C  &     44.4      &  71.2       &  15.66  &  2.84   &  1.60                        \\
$^{16}$O  &     59.2      &  106.3      &  13.26  &  4.47   &  1.80                        \\
$^{17}$O  &     62.7      &  106.6      &  13.35  &  5.87   &  1.70                        \\
$^{18}$O  &     65.8      &  117.7      &  12.51  &  5.26   &  1.79                        \\
$^{20}$Ne &     74.0      &  149.5      &  11.88  &  6.23   &  2.02                        \\
$^{24}$Mg &     88.8      &  128.0      &  12.62  &  7.04       &  1.44                        \\
$^{36}$Ar &    133.2      &  235.5      &  11.05  &  12.05  &  1.78                        \\
$^{40}$Ca &    148.0      &  274.8      &  10.77  &  13.74  &  1.86                        \\
\hline \hline
\end{tabular}
\end{center}
\end{table}


When the particle-hole interaction is turned on, the  1p-1h states mix and are pushed up in energy forming the collective dipole resonance.
In our 1p-1h model the summed $B(E1)$ strength remains the same after mixing.
The resulting {\small SM} energy-weighted sum over $E1$ excitations up to $E_{_{max}}^{^{SM}}$ is then given by,
\begin{eqnarray}
S(E1)^{^{SM}}=\sum\limits_{n}^{E_{_{max}}^{^{SM}}}  E_n^{SM} ~ B(E1_n)^{SM}.
\label{eq:SE1wbp}
\end{eqnarray}
Our results for $S(E1)^{^{SM}}$ are presented in Table~\ref{tab:BE1s}, together with the fraction of the {\small TRK} sum rule
exhausted for each nucleus,
\begin{equation}
f_{_{GDR}} = \frac{S(E1)^{^{SM}}}{{S}(E1)^{^{TRK}}}.
\end{equation}
The $\sum\limits_{n} {B(E1_n)^{SM}}$ values are presented in
Table~\ref{tab:tab1}.
In general, $f_{_{GDR}}$ values exceedingly exhaust the {\small TRK} sum rule
by approximately 1.5-2 times.
In $^{16}$O, the  1p-1h  states are formed by linear
combinations of the five possible excitations. The strongest
state that comes at 23.6 MeV contains 89\% of the $E1$ strength.
The {\small TRK} sum is enhanced by a factor of 23.6/13.3 = 1.8.
For $^{24}$Mg,   we are missing some of the $E1$ strength in the lowest 2000 1$^-$ $T=1$ states, presenting the smallest $f_{_{GDR}}$ value of 1.44.

Below $^{16}$O one includes both $  0s  $ to $  0p  $ and $  0p  $ to $  0d-1s  $.
Above $^{16}$O and below $^{40}$Ca one includes $  0p  $ to $  0d-1s  $ and $  1d-0s  $ to $  0f-1p  $.
For these nuclei the strength becomes more fragmented --- as shown in Fig.~\ref{fig:20Ne24Mg36Ar}\footnote{
Such a broad and fragmented $B(E1; 0^+_{_1} \rightarrow 1^-_{_{GDR}})$ distribution for $^{24}$Mg is supported by high-resolution measurements of the {\small GDR} region~\cite{fearick2018origin}, where the fine structure is expected to arise from the deformation driven by $\alpha$ clustering. Albeit high-resolution data being also scarce,  promising
zero-degree $(p,p^{\prime})$ measurements at the required proton energies of $\gtrapprox$200 MeV may soon become available for $A<60$ nuclei through the {\small PANDORA} project~\cite{tamii2022pandora}.} --- due to coupling with 1$^{-}$ states that are built on other other
(non closed-shell) positive-parity states. For example, in $^{20}$Ne there are 1525 non-spurious 1$^{-}$, $  T=1  $ states
(one can make spurious 1$^{-}$, $  T=1  $ states by coupling
spurious 1$^{-}$, $  T=0  $ states to positive parity states with $  T=1  $). As expected, in $^{20}$Ne there is a negligible $E1$ strength at low
excitation energies below $\approx 14$ MeV, in agreement with recent calculations  using the microscopic configuration interaction method~\cite{sieja2023brink}. The same happens for $^{24}$Mg. Nonetheless, some $E1$ strength starts developing at low energies in $^{36}$Ar as the $Z=N=20$ shell closures  approach.

The {\small TRK} sum is increased due to the increase in energy
of the {\small GDR} from 1$\hbar\omega$. In these  {\small SM} calculations,
the  {\small TRK} enhancement factor comes from using realistic
single-particle energies (as given above for $^{16}$O)  and the attractive particle-hole interaction, which is dominated by the central part of the potential.

The centroid of our calculated {\small GDR}  is systematically higher that those obtained from $(p,p^{\prime})$ experiments for $sd$ shell nuclei~\cite{fearick2018origin}.
For example, for $^{24}$Mg  the experimental {\small GDR} is distributed over a range of 16-18 MeV~\cite{fearick2018origin} compared to the calculations in Fig.~\ref{fig:20Ne24Mg36Ar}
that have a range of 16-26 MeV. Using an experimental centroid energy of 17 MeV together with our $B(E1)$ value would give $f_{_{GDR}}\approx 1.35$.
The data used to obtained the parameters of the {\small WBP} and {\small FSU} Hamiltonians
did not include the energy of the {\small GDR}.
Thus, it is possible that the energy of
the calculated energies of the {\small GDR} states could be modified by
further adjustments of the potential and {\small TBME}
parameters of the {\small FSU} Hamiltonian.
 It is also possible
that the energy of the {\small GDR} is lowered
and the width is increased by coupling with
the continuum, which is not included in the calculations.
Values of $f_{_{GDR}}\gg1$  have generally been associated with
additional degrees of freedom arising from velocity-dependent interactions or short-range correlations --- e.g. the aforementioned cluster formations --- between protons and neutrons~\cite{levinger1950dipole,ferentz1953giant,johnson1955classical,weisskopf1957problem,rand1957appreciation,migdal1965theory,brenig1965advances}.

\section{Conclusions and Further Work}

This work explores the {\small GDR} and related electric dipole polarizability effects from {\small SM} calculations using, for the first time, the {\small WBP} and {\small FSU} interactions,
and extend previous knowledge towards the end of the {\it  sd} shell.
Strong deviations from the hydrodynamic model are calculated following the smooth trend  predicted independently by the
leptodermus approximation using the {\small FRDM} model, and in agreement with calculations from first principles at shell closures. The sum of $E1$ strengths also follow a smooth, linear trend directly proportional to the mass number,
which supports to some extent the complementary macroscopic nature of the {\small GDR}.
Throughout, the classical oscillator strength sum rule is  exceedingly exhausted by a factor of 1.5-2.
The enhancement of the energy-weighted sum over $E1$ excitations with respect to the {\small TRK}
sum rule comes from  the use of realistic single-particle energies and the repulsive particle-hole interaction.
The inclusion of  missing admixtures with $1\hbar\omega$ 1p-1h + 2p-2h transitions may slightly suppress the $E1$ strength in closer agreement with photo-absorption cross-section measurements.

Further {\small NCSM} calculations of $sd$-shell nuclei up to $A=24$  are now potentially available~\cite{sarma2022ab} and their results keenly expected; particularly using a new generation of $\chi EFT$ interactions and large $N_{max}$ basis sizes~\cite{sarma2022ab}.
Considerable experimental work is also required to accurately determine the currently scarce $(\gamma,p)$ cross sections as well as  the high-resolution structure of the {\small GDR}
in order to
benchmark  the calculations presented here.
More broadly, it will also be exciting to investigate the $E1$ polarizability of excited states~\cite{eichler1964second} as well as unstable nuclei, where $\kappa(g.s.)=1.26(10)$ remains the only data point corresponding to the ground state of the semi-magic nucleus $^{68}$Ni~\cite{rossi2013measurement}.


BAB acknowledges NSF grant PHY-2110365.

%

\bibliographystyle{apsrev4-2}

\bibliography{polags_Sep23}

\begin{thebibliography}{129}%
\makeatletter
\providecommand \@ifxundefined [1]{%
 \@ifx{#1\undefined}
}%
\providecommand \@ifnum [1]{%
 \ifnum #1\expandafter \@firstoftwo
 \else \expandafter \@secondoftwo
 \fi
}%
\providecommand \@ifx [1]{%
 \ifx #1\expandafter \@firstoftwo
 \else \expandafter \@secondoftwo
 \fi
}%
\providecommand \natexlab [1]{#1}%
\providecommand \enquote  [1]{``#1''}%
\providecommand \bibnamefont  [1]{#1}%
\providecommand \bibfnamefont [1]{#1}%
\providecommand \citenamefont [1]{#1}%
\providecommand \href@noop [0]{\@secondoftwo}%
\providecommand \href [0]{\begingroup \@sanitize@url \@href}%
\providecommand \@href[1]{\@@startlink{#1}\@@href}%
\providecommand \@@href[1]{\endgroup#1\@@endlink}%
\providecommand \@sanitize@url [0]{\catcode `\\12\catcode `\$12\catcode
  `\&12\catcode `\#12\catcode `\^12\catcode `\_12\catcode `\%12\relax}%
\providecommand \@@startlink[1]{}%
\providecommand \@@endlink[0]{}%
\providecommand \url  [0]{\begingroup\@sanitize@url \@url }%
\providecommand \@url [1]{\endgroup\@href {#1}{\urlprefix }}%
\providecommand \urlprefix  [0]{URL }%
\providecommand \Eprint [0]{\href }%
\providecommand \doibase [0]{https://doi.org/}%
\providecommand \selectlanguage [0]{\@gobble}%
\providecommand \bibinfo  [0]{\@secondoftwo}%
\providecommand \bibfield  [0]{\@secondoftwo}%
\providecommand \translation [1]{[#1]}%
\providecommand \BibitemOpen [0]{}%
\providecommand \bibitemStop [0]{}%
\providecommand \bibitemNoStop [0]{.\EOS\space}%
\providecommand \EOS [0]{\spacefactor3000\relax}%
\providecommand \BibitemShut  [1]{\csname bibitem#1\endcsname}%
\let\auto@bib@innerbib\@empty
\bibitem [{\citenamefont {Baldwin}\ and\ \citenamefont
  {Klaiber}(1947)}]{baldwin1947photo}%
  \BibitemOpen
  \bibfield  {author} {\bibinfo {author} {\bibfnamefont {G.~C.}\ \bibnamefont
  {Baldwin}}\ and\ \bibinfo {author} {\bibfnamefont {G.~S.}\ \bibnamefont
  {Klaiber}},\ }\href@noop {} {\bibfield  {journal} {\bibinfo  {journal}
  {Physical Review}\ }\textbf {\bibinfo {volume} {71}},\ \bibinfo {pages} {3}
  (\bibinfo {year} {1947})}\BibitemShut {NoStop}%
\bibitem [{\citenamefont {Migdal}(1945)}]{migdal1945quadrupole}%
  \BibitemOpen
  \bibfield  {author} {\bibinfo {author} {\bibfnamefont {A.}~\bibnamefont
  {Migdal}},\ }\href@noop {} {\bibfield  {journal} {\bibinfo  {journal}
  {Zhurnal Eksperimentalnoi i Teoreticheskoi Fiziki}\ }\textbf {\bibinfo
  {volume} {15}},\ \bibinfo {pages} {81} (\bibinfo {year} {1945})}\BibitemShut
  {NoStop}%
\bibitem [{\citenamefont {Goldhaber}\ and\ \citenamefont
  {Teller}(1948)}]{goldhaber1948nuclear}%
  \BibitemOpen
  \bibfield  {author} {\bibinfo {author} {\bibfnamefont {M.}~\bibnamefont
  {Goldhaber}}\ and\ \bibinfo {author} {\bibfnamefont {E.}~\bibnamefont
  {Teller}},\ }\href@noop {} {\bibfield  {journal} {\bibinfo  {journal}
  {Physical Review}\ }\textbf {\bibinfo {volume} {74}},\ \bibinfo {pages}
  {1046} (\bibinfo {year} {1948})}\BibitemShut {NoStop}%
\bibitem [{\citenamefont {Steinwedel}\ \emph {et~al.}(1950)\citenamefont
  {Steinwedel}, \citenamefont {Jensen},\ and\ \citenamefont
  {Jensen}}]{steinwedel1950nuclear}%
  \BibitemOpen
  \bibfield  {author} {\bibinfo {author} {\bibfnamefont {H.}~\bibnamefont
  {Steinwedel}}, \bibinfo {author} {\bibfnamefont {J.~H.~D.}\ \bibnamefont
  {Jensen}},\ and\ \bibinfo {author} {\bibfnamefont {P.}~\bibnamefont
  {Jensen}},\ }\href@noop {} {\bibfield  {journal} {\bibinfo  {journal}
  {Physical Review}\ }\textbf {\bibinfo {volume} {79}},\ \bibinfo {pages}
  {1019} (\bibinfo {year} {1950})}\BibitemShut {NoStop}%
\bibitem [{\citenamefont {Levinger}\ and\ \citenamefont
  {Kent}(1954)}]{levinger1954independent}%
  \BibitemOpen
  \bibfield  {author} {\bibinfo {author} {\bibfnamefont {J.~S.}\ \bibnamefont
  {Levinger}}\ and\ \bibinfo {author} {\bibfnamefont {D.~C.}\ \bibnamefont
  {Kent}},\ }\href@noop {} {\bibfield  {journal} {\bibinfo  {journal} {Physical
  Review}\ }\textbf {\bibinfo {volume} {95}},\ \bibinfo {pages} {418} (\bibinfo
  {year} {1954})}\BibitemShut {NoStop}%
\bibitem [{\citenamefont {Wilkinson}(1956)}]{wilkinson1956nuclear}%
  \BibitemOpen
  \bibfield  {author} {\bibinfo {author} {\bibfnamefont {D.~H.}\ \bibnamefont
  {Wilkinson}},\ }\href@noop {} {\bibfield  {journal} {\bibinfo  {journal}
  {Physica}\ }\textbf {\bibinfo {volume} {22}},\ \bibinfo {pages} {1039}
  (\bibinfo {year} {1956})}\BibitemShut {NoStop}%
\bibitem [{\citenamefont {Balashov}(1962)}]{balashov1962relation}%
  \BibitemOpen
  \bibfield  {author} {\bibinfo {author} {\bibfnamefont {V.~V.}\ \bibnamefont
  {Balashov}},\ }\href@noop {} {\bibfield  {journal} {\bibinfo  {journal}
  {Zhurnal {\^E}ksperimental'noi i Teoreticheskoi Fiziki}\ }\textbf {\bibinfo
  {volume} {42}},\ \bibinfo {pages} {275} (\bibinfo {year} {1962})}\BibitemShut
  {NoStop}%
\bibitem [{\citenamefont {Danos}\ and\ \citenamefont
  {Fuller}(1965)}]{danos1965photonuclear}%
  \BibitemOpen
  \bibfield  {author} {\bibinfo {author} {\bibfnamefont {M.}~\bibnamefont
  {Danos}}\ and\ \bibinfo {author} {\bibfnamefont {E.~G.}\ \bibnamefont
  {Fuller}},\ }\href@noop {} {\bibfield  {journal} {\bibinfo  {journal} {Annual
  Review of Nuclear Science}\ }\textbf {\bibinfo {volume} {15}},\ \bibinfo
  {pages} {29} (\bibinfo {year} {1965})}\BibitemShut {NoStop}%
\bibitem [{\citenamefont {Weizs{\"a}cker}(1935)}]{weizsacker1935theorie}%
  \BibitemOpen
  \bibfield  {author} {\bibinfo {author} {\bibfnamefont {C.~F.}\ \bibnamefont
  {Weizs{\"a}cker}},\ }\href@noop {} {\bibfield  {journal} {\bibinfo  {journal}
  {Zeitschrift f{\"u}r Physik}\ }\textbf {\bibinfo {volume} {96}},\ \bibinfo
  {pages} {431} (\bibinfo {year} {1935})}\BibitemShut {NoStop}%
\bibitem [{\citenamefont {Bethe}\ and\ \citenamefont
  {Bacher}(1936)}]{bethe1936nuclear}%
  \BibitemOpen
  \bibfield  {author} {\bibinfo {author} {\bibfnamefont {H.~A.}\ \bibnamefont
  {Bethe}}\ and\ \bibinfo {author} {\bibfnamefont {R.~F.}\ \bibnamefont
  {Bacher}},\ }\href@noop {} {\bibfield  {journal} {\bibinfo  {journal}
  {Reviews of Modern Physics}\ }\textbf {\bibinfo {volume} {8}},\ \bibinfo
  {pages} {82} (\bibinfo {year} {1936})}\BibitemShut {NoStop}%
\bibitem [{\citenamefont {Berman}\ and\ \citenamefont
  {Fultz}(1975)}]{berman1975measurements}%
  \BibitemOpen
  \bibfield  {author} {\bibinfo {author} {\bibfnamefont {B.~L.}\ \bibnamefont
  {Berman}}\ and\ \bibinfo {author} {\bibfnamefont {S.}~\bibnamefont {Fultz}},\
  }\href@noop {} {\bibfield  {journal} {\bibinfo  {journal} {Reviews of Modern
  Physics}\ }\textbf {\bibinfo {volume} {47}},\ \bibinfo {pages} {713}
  (\bibinfo {year} {1975})}\BibitemShut {NoStop}%
\bibitem [{\citenamefont {Levinger}(1957)}]{levinger1957migdal}%
  \BibitemOpen
  \bibfield  {author} {\bibinfo {author} {\bibfnamefont {J.}~\bibnamefont
  {Levinger}},\ }\href@noop {} {\bibfield  {journal} {\bibinfo  {journal}
  {Physical Review}\ }\textbf {\bibinfo {volume} {107}},\ \bibinfo {pages}
  {554} (\bibinfo {year} {1957})}\BibitemShut {NoStop}%
\bibitem [{\citenamefont {Flambaum}\ \emph {et~al.}(2021)\citenamefont
  {Flambaum}, \citenamefont {Samsonov}, \citenamefont {Tan},\ and\
  \citenamefont {Viatkina}}]{flambaum2021nuclear}%
  \BibitemOpen
  \bibfield  {author} {\bibinfo {author} {\bibfnamefont {V.}~\bibnamefont
  {Flambaum}}, \bibinfo {author} {\bibfnamefont {I.}~\bibnamefont {Samsonov}},
  \bibinfo {author} {\bibfnamefont {H.~T.}\ \bibnamefont {Tan}},\ and\ \bibinfo
  {author} {\bibfnamefont {A.}~\bibnamefont {Viatkina}},\ }\href@noop {}
  {\bibfield  {journal} {\bibinfo  {journal} {Physical Review A}\ }\textbf
  {\bibinfo {volume} {103}},\ \bibinfo {pages} {032811} (\bibinfo {year}
  {2021})}\BibitemShut {NoStop}%
\bibitem [{\citenamefont {de~Boer}\ and\ \citenamefont
  {Eichler}(1968)}]{deBoer1968}%
  \BibitemOpen
  \bibfield  {author} {\bibinfo {author} {\bibfnamefont {J.}~\bibnamefont
  {de~Boer}}\ and\ \bibinfo {author} {\bibfnamefont {J.}~\bibnamefont
  {Eichler}},\ }\href@noop {} {\bibfield  {journal} {\bibinfo  {journal}
  {Advances in Nuclear Physics}\ }\textbf {\bibinfo {volume} {1}},\ \bibinfo
  {pages} {1} (\bibinfo {year} {1968})}\BibitemShut {NoStop}%
\bibitem [{\citenamefont {Rose}(1957)}]{rose1957elementary}%
  \BibitemOpen
  \bibfield  {author} {\bibinfo {author} {\bibfnamefont {M.}~\bibnamefont
  {Rose}},\ }\href@noop {} {\bibfield  {journal} {\bibinfo  {journal} {New
  York}\ } (\bibinfo {year} {1957})}\BibitemShut {NoStop}%
\bibitem [{\citenamefont {Messiah}(1961)}]{messiah1961quantum}%
  \BibitemOpen
  \bibfield  {author} {\bibinfo {author} {\bibfnamefont {A.}~\bibnamefont
  {Messiah}},\ }\href@noop {} {\bibinfo {title} {Quantum mechanics vol 1 \& 2
  tr. gm temmer}} (\bibinfo {year} {1961})\BibitemShut {NoStop}%
\bibitem [{\citenamefont {Levinger}(1960)}]{levinger1960}%
  \BibitemOpen
  \bibfield  {author} {\bibinfo {author} {\bibfnamefont {J.~S.}\ \bibnamefont
  {Levinger}},\ }\href@noop {} {\emph {\bibinfo {title} {Nuclear
  Photo-Disintegration}}}\ (\bibinfo  {publisher} {Oxford University Press},\
  \bibinfo {address} {Oxford},\ \bibinfo {year} {1960})\BibitemShut {NoStop}%
\bibitem [{\citenamefont {Migdal}\ \emph {et~al.}(1965)\citenamefont {Migdal},
  \citenamefont {Lushnikov},\ and\ \citenamefont
  {Zaretsky}}]{migdal1965theory}%
  \BibitemOpen
  \bibfield  {author} {\bibinfo {author} {\bibfnamefont {A.}~\bibnamefont
  {Migdal}}, \bibinfo {author} {\bibfnamefont {A.}~\bibnamefont {Lushnikov}},\
  and\ \bibinfo {author} {\bibfnamefont {D.}~\bibnamefont {Zaretsky}},\
  }\href@noop {} {\bibfield  {journal} {\bibinfo  {journal} {Nuclear Physics}\
  }\textbf {\bibinfo {volume} {66}},\ \bibinfo {pages} {193} (\bibinfo {year}
  {1965})}\BibitemShut {NoStop}%
\bibitem [{\citenamefont {Dietrich}\ and\ \citenamefont
  {Berman}(1988)}]{dietrich1988atlas}%
  \BibitemOpen
  \bibfield  {author} {\bibinfo {author} {\bibfnamefont {S.~S.}\ \bibnamefont
  {Dietrich}}\ and\ \bibinfo {author} {\bibfnamefont {B.~L.}\ \bibnamefont
  {Berman}},\ }\href@noop {} {\bibfield  {journal} {\bibinfo  {journal} {Atomic
  Data and Nuclear Data Tables}\ }\textbf {\bibinfo {volume} {38}},\ \bibinfo
  {pages} {199} (\bibinfo {year} {1988})}\BibitemShut {NoStop}%
\bibitem [{\citenamefont {Ahrens}\ \emph {et~al.}(1976)\citenamefont {Ahrens},
  \citenamefont {Gimm}, \citenamefont {Zieger},\ and\ \citenamefont
  {Ziegler}}]{ahrens1976experimental}%
  \BibitemOpen
  \bibfield  {author} {\bibinfo {author} {\bibfnamefont {J.}~\bibnamefont
  {Ahrens}}, \bibinfo {author} {\bibfnamefont {H.}~\bibnamefont {Gimm}},
  \bibinfo {author} {\bibfnamefont {A.}~\bibnamefont {Zieger}},\ and\ \bibinfo
  {author} {\bibfnamefont {B.}~\bibnamefont {Ziegler}},\ }\href@noop {}
  {\bibfield  {journal} {\bibinfo  {journal} {Il Nuovo Cimento A (1965-1970)}\
  }\textbf {\bibinfo {volume} {32}},\ \bibinfo {pages} {364} (\bibinfo {year}
  {1976})}\BibitemShut {NoStop}%
\bibitem [{\citenamefont {Knupfer}\ and\ \citenamefont
  {Richter}(1985)}]{knupfer1985scaling}%
  \BibitemOpen
  \bibfield  {author} {\bibinfo {author} {\bibfnamefont {W.}~\bibnamefont
  {Knupfer}}\ and\ \bibinfo {author} {\bibfnamefont {A.}~\bibnamefont
  {Richter}},\ }\href@noop {} {\bibfield  {journal} {\bibinfo  {journal}
  {Zeitschrift fur Physik A Atoms and Nuclei}\ }\textbf {\bibinfo {volume}
  {320}},\ \bibinfo {pages} {253} (\bibinfo {year} {1985})}\BibitemShut
  {NoStop}%
\bibitem [{\citenamefont {Knupfer}\ and\ \citenamefont
  {Richter}(1981)}]{knupfer1981effect}%
  \BibitemOpen
  \bibfield  {author} {\bibinfo {author} {\bibfnamefont {W.}~\bibnamefont
  {Knupfer}}\ and\ \bibinfo {author} {\bibfnamefont {A.}~\bibnamefont
  {Richter}},\ }\href@noop {} {\bibfield  {journal} {\bibinfo  {journal}
  {Physics Letters B}\ }\textbf {\bibinfo {volume} {107}},\ \bibinfo {pages}
  {325} (\bibinfo {year} {1981})}\BibitemShut {NoStop}%
\bibitem [{\citenamefont {Plujko}\ \emph {et~al.}(2018)\citenamefont {Plujko},
  \citenamefont {Gorbachenko}, \citenamefont {Capote},\ and\ \citenamefont
  {Dimitriou}}]{plujko2018giant}%
  \BibitemOpen
  \bibfield  {author} {\bibinfo {author} {\bibfnamefont {V.~A.}\ \bibnamefont
  {Plujko}}, \bibinfo {author} {\bibfnamefont {O.~M.}\ \bibnamefont
  {Gorbachenko}}, \bibinfo {author} {\bibfnamefont {R.}~\bibnamefont
  {Capote}},\ and\ \bibinfo {author} {\bibfnamefont {P.}~\bibnamefont
  {Dimitriou}},\ }\href@noop {} {\bibfield  {journal} {\bibinfo  {journal}
  {Atomic Data and Nuclear Data Tables}\ }\textbf {\bibinfo {volume} {123}},\
  \bibinfo {pages} {1} (\bibinfo {year} {2018})}\BibitemShut {NoStop}%
\bibitem [{\citenamefont {Kawano}\ \emph {et~al.}(2020)\citenamefont {Kawano},
  \citenamefont {Cho}, \citenamefont {Dimitriou}, \citenamefont {Filipescu},
  \citenamefont {Iwamoto}, \citenamefont {Plujko}, \citenamefont {Tao},
  \citenamefont {Utsunomiya}, \citenamefont {Varlamov}, \citenamefont {Xu}
  \emph {et~al.}}]{kawano2020iaea}%
  \BibitemOpen
  \bibfield  {author} {\bibinfo {author} {\bibfnamefont {T.}~\bibnamefont
  {Kawano}}, \bibinfo {author} {\bibfnamefont {Y.}~\bibnamefont {Cho}},
  \bibinfo {author} {\bibfnamefont {P.}~\bibnamefont {Dimitriou}}, \bibinfo
  {author} {\bibfnamefont {D.}~\bibnamefont {Filipescu}}, \bibinfo {author}
  {\bibfnamefont {N.}~\bibnamefont {Iwamoto}}, \bibinfo {author} {\bibfnamefont
  {V.}~\bibnamefont {Plujko}}, \bibinfo {author} {\bibfnamefont
  {X.}~\bibnamefont {Tao}}, \bibinfo {author} {\bibfnamefont {H.}~\bibnamefont
  {Utsunomiya}}, \bibinfo {author} {\bibfnamefont {V.}~\bibnamefont
  {Varlamov}}, \bibinfo {author} {\bibfnamefont {R.}~\bibnamefont {Xu}}, \emph
  {et~al.},\ }\href@noop {} {\bibfield  {journal} {\bibinfo  {journal} {Nuclear
  Data Sheets}\ }\textbf {\bibinfo {volume} {163}},\ \bibinfo {pages} {109}
  (\bibinfo {year} {2020})}\BibitemShut {NoStop}%
\bibitem [{\citenamefont {Ishkhanov}\ and\ \citenamefont
  {Kapitonov}(2021)}]{ishkhanov2021giant}%
  \BibitemOpen
  \bibfield  {author} {\bibinfo {author} {\bibfnamefont {B.~S.}\ \bibnamefont
  {Ishkhanov}}\ and\ \bibinfo {author} {\bibfnamefont {I.~M.}\ \bibnamefont
  {Kapitonov}},\ }\href@noop {} {\bibfield  {journal} {\bibinfo  {journal}
  {Physics-Uspekhi}\ }\textbf {\bibinfo {volume} {64}},\ \bibinfo {pages} {141}
  (\bibinfo {year} {2021})}\BibitemShut {NoStop}%
\bibitem [{\citenamefont {Orce}(2022)}]{orce2022competition}%
  \BibitemOpen
  \bibfield  {author} {\bibinfo {author} {\bibfnamefont {J.~N.}\ \bibnamefont
  {Orce}},\ }\href@noop {} {\bibfield  {journal} {\bibinfo  {journal} {Atomic
  Data and Nuclear Data Tables}\ }\textbf {\bibinfo {volume} {145}},\ \bibinfo
  {pages} {101511} (\bibinfo {year} {2022})}\BibitemShut {NoStop}%
\bibitem [{\citenamefont {Rossi}\ \emph {et~al.}(2013)\citenamefont {Rossi},
  \citenamefont {Adrich}, \citenamefont {Aksouh}, \citenamefont {Alvarez-Pol},
  \citenamefont {Aumann}, \citenamefont {Benlliure}, \citenamefont
  {B{\"o}hmer}, \citenamefont {Boretzky}, \citenamefont {Casarejos},
  \citenamefont {Chartier} \emph {et~al.}}]{rossi2013measurement}%
  \BibitemOpen
  \bibfield  {author} {\bibinfo {author} {\bibfnamefont {D.~M.}\ \bibnamefont
  {Rossi}}, \bibinfo {author} {\bibfnamefont {P.}~\bibnamefont {Adrich}},
  \bibinfo {author} {\bibfnamefont {F.}~\bibnamefont {Aksouh}}, \bibinfo
  {author} {\bibfnamefont {H.}~\bibnamefont {Alvarez-Pol}}, \bibinfo {author}
  {\bibfnamefont {T.}~\bibnamefont {Aumann}}, \bibinfo {author} {\bibfnamefont
  {J.}~\bibnamefont {Benlliure}}, \bibinfo {author} {\bibfnamefont
  {M.}~\bibnamefont {B{\"o}hmer}}, \bibinfo {author} {\bibfnamefont
  {K.}~\bibnamefont {Boretzky}}, \bibinfo {author} {\bibfnamefont
  {E.}~\bibnamefont {Casarejos}}, \bibinfo {author} {\bibfnamefont
  {M.}~\bibnamefont {Chartier}}, \emph {et~al.},\ }\href@noop {} {\bibfield
  {journal} {\bibinfo  {journal} {Physical Review Letters}\ }\textbf {\bibinfo
  {volume} {111}},\ \bibinfo {pages} {242503} (\bibinfo {year}
  {2013})}\BibitemShut {NoStop}%
\bibitem [{\citenamefont {Tamii}\ \emph {et~al.}(2011)\citenamefont {Tamii},
  \citenamefont {Poltoratska}, \citenamefont {von Neumann-Cosel}, \citenamefont
  {Fujita}, \citenamefont {Adachi}, \citenamefont {Bertulani}, \citenamefont
  {Carter}, \citenamefont {Dozono}, \citenamefont {Fujita}, \citenamefont
  {Fujita} \emph {et~al.}}]{tamii2011complete}%
  \BibitemOpen
  \bibfield  {author} {\bibinfo {author} {\bibfnamefont {A.}~\bibnamefont
  {Tamii}}, \bibinfo {author} {\bibfnamefont {I.}~\bibnamefont {Poltoratska}},
  \bibinfo {author} {\bibfnamefont {P.}~\bibnamefont {von Neumann-Cosel}},
  \bibinfo {author} {\bibfnamefont {Y.}~\bibnamefont {Fujita}}, \bibinfo
  {author} {\bibfnamefont {T.}~\bibnamefont {Adachi}}, \bibinfo {author}
  {\bibfnamefont {C.}~\bibnamefont {Bertulani}}, \bibinfo {author}
  {\bibfnamefont {J.}~\bibnamefont {Carter}}, \bibinfo {author} {\bibfnamefont
  {M.}~\bibnamefont {Dozono}}, \bibinfo {author} {\bibfnamefont
  {H.}~\bibnamefont {Fujita}}, \bibinfo {author} {\bibfnamefont
  {K.}~\bibnamefont {Fujita}}, \emph {et~al.},\ }\href@noop {} {\bibfield
  {journal} {\bibinfo  {journal} {Physical Review Letters}\ }\textbf {\bibinfo
  {volume} {107}},\ \bibinfo {pages} {062502} (\bibinfo {year}
  {2011})}\BibitemShut {NoStop}%
\bibitem [{\citenamefont {Roca-Maza}\ \emph {et~al.}(2015)\citenamefont
  {Roca-Maza}, \citenamefont {Vi{\~n}as}, \citenamefont {Centelles},
  \citenamefont {Agrawal}, \citenamefont {Colo}, \citenamefont {Paar},
  \citenamefont {Piekarewicz},\ and\ \citenamefont
  {Vretenar}}]{roca2015neutron}%
  \BibitemOpen
  \bibfield  {author} {\bibinfo {author} {\bibfnamefont {X.}~\bibnamefont
  {Roca-Maza}}, \bibinfo {author} {\bibfnamefont {X.}~\bibnamefont
  {Vi{\~n}as}}, \bibinfo {author} {\bibfnamefont {M.}~\bibnamefont
  {Centelles}}, \bibinfo {author} {\bibfnamefont {B.~K.}\ \bibnamefont
  {Agrawal}}, \bibinfo {author} {\bibfnamefont {G.}~\bibnamefont {Colo}},
  \bibinfo {author} {\bibfnamefont {N.}~\bibnamefont {Paar}}, \bibinfo {author}
  {\bibfnamefont {J.}~\bibnamefont {Piekarewicz}},\ and\ \bibinfo {author}
  {\bibfnamefont {D.}~\bibnamefont {Vretenar}},\ }\href@noop {} {\bibfield
  {journal} {\bibinfo  {journal} {Physical Review C}\ }\textbf {\bibinfo
  {volume} {92}},\ \bibinfo {pages} {064304} (\bibinfo {year}
  {2015})}\BibitemShut {NoStop}%
\bibitem [{\citenamefont {Hashimoto}\ \emph {et~al.}(2015)\citenamefont
  {Hashimoto}, \citenamefont {Krumbholz}, \citenamefont {Reinhard},
  \citenamefont {Tamii}, \citenamefont {von Neumann-Cosel}, \citenamefont
  {Adachi}, \citenamefont {Aoi}, \citenamefont {Bertulani}, \citenamefont
  {Fujita}, \citenamefont {Fujita} \emph {et~al.}}]{hashimoto2015dipole}%
  \BibitemOpen
  \bibfield  {author} {\bibinfo {author} {\bibfnamefont {T.}~\bibnamefont
  {Hashimoto}}, \bibinfo {author} {\bibfnamefont {A.~M.}\ \bibnamefont
  {Krumbholz}}, \bibinfo {author} {\bibfnamefont {P.~G.}\ \bibnamefont
  {Reinhard}}, \bibinfo {author} {\bibfnamefont {A.}~\bibnamefont {Tamii}},
  \bibinfo {author} {\bibfnamefont {P.}~\bibnamefont {von Neumann-Cosel}},
  \bibinfo {author} {\bibfnamefont {T.}~\bibnamefont {Adachi}}, \bibinfo
  {author} {\bibfnamefont {N.}~\bibnamefont {Aoi}}, \bibinfo {author}
  {\bibfnamefont {C.~A.}\ \bibnamefont {Bertulani}}, \bibinfo {author}
  {\bibfnamefont {H.}~\bibnamefont {Fujita}}, \bibinfo {author} {\bibfnamefont
  {Y.}~\bibnamefont {Fujita}}, \emph {et~al.},\ }\href@noop {} {\bibfield
  {journal} {\bibinfo  {journal} {Physical Review C}\ }\textbf {\bibinfo
  {volume} {92}},\ \bibinfo {pages} {031305} (\bibinfo {year}
  {2015})}\BibitemShut {NoStop}%
\bibitem [{\citenamefont {Roca-Maza}\ and\ \citenamefont
  {Paar}(2018)}]{roca2018nuclear}%
  \BibitemOpen
  \bibfield  {author} {\bibinfo {author} {\bibfnamefont {X.}~\bibnamefont
  {Roca-Maza}}\ and\ \bibinfo {author} {\bibfnamefont {N.}~\bibnamefont
  {Paar}},\ }\href@noop {} {\bibfield  {journal} {\bibinfo  {journal} {Progress
  in Particle and Nuclear Physics}\ }\textbf {\bibinfo {volume} {101}},\
  \bibinfo {pages} {96} (\bibinfo {year} {2018})}\BibitemShut {NoStop}%
\bibitem [{\citenamefont {Bassauer}\ \emph {et~al.}(2020)\citenamefont
  {Bassauer}, \citenamefont {von Neumann-Cosel}, \citenamefont {Reinhard},
  \citenamefont {Tamii}, \citenamefont {Adachi}, \citenamefont {Bertulani},
  \citenamefont {Chan}, \citenamefont {Col{\`o}}, \citenamefont {D'Alessio},
  \citenamefont {Fujioka} \emph {et~al.}}]{bassauer2020evolution}%
  \BibitemOpen
  \bibfield  {author} {\bibinfo {author} {\bibfnamefont {S.}~\bibnamefont
  {Bassauer}}, \bibinfo {author} {\bibfnamefont {P.}~\bibnamefont {von
  Neumann-Cosel}}, \bibinfo {author} {\bibfnamefont {P.-G.}\ \bibnamefont
  {Reinhard}}, \bibinfo {author} {\bibfnamefont {A.}~\bibnamefont {Tamii}},
  \bibinfo {author} {\bibfnamefont {S.}~\bibnamefont {Adachi}}, \bibinfo
  {author} {\bibfnamefont {C.~A.}\ \bibnamefont {Bertulani}}, \bibinfo {author}
  {\bibfnamefont {P.}~\bibnamefont {Chan}}, \bibinfo {author} {\bibfnamefont
  {G.}~\bibnamefont {Col{\`o}}}, \bibinfo {author} {\bibfnamefont
  {A.}~\bibnamefont {D'Alessio}}, \bibinfo {author} {\bibfnamefont
  {H.}~\bibnamefont {Fujioka}}, \emph {et~al.},\ }\href@noop {} {\bibfield
  {journal} {\bibinfo  {journal} {Physics Letters B}\ }\textbf {\bibinfo
  {volume} {810}},\ \bibinfo {pages} {135804} (\bibinfo {year}
  {2020})}\BibitemShut {NoStop}%
\bibitem [{\citenamefont {Orce}(2020)}]{orce2020polarizability}%
  \BibitemOpen
  \bibfield  {author} {\bibinfo {author} {\bibfnamefont {J.~N.}\ \bibnamefont
  {Orce}},\ }\href@noop {} {\bibfield  {journal} {\bibinfo  {journal}
  {International Journal of Modern Physics E}\ }\textbf {\bibinfo {volume}
  {29}},\ \bibinfo {pages} {2030002} (\bibinfo {year} {2020})}\BibitemShut
  {NoStop}%
\bibitem [{\citenamefont {Nathans}\ and\ \citenamefont
  {Halpern}(1953)}]{nathans1953excitation}%
  \BibitemOpen
  \bibfield  {author} {\bibinfo {author} {\bibfnamefont {R.}~\bibnamefont
  {Nathans}}\ and\ \bibinfo {author} {\bibfnamefont {J.}~\bibnamefont
  {Halpern}},\ }\href@noop {} {\bibfield  {journal} {\bibinfo  {journal}
  {Physical Review}\ }\textbf {\bibinfo {volume} {92}},\ \bibinfo {pages} {940}
  (\bibinfo {year} {1953})}\BibitemShut {NoStop}%
\bibitem [{\citenamefont {Denisov}\ and\ \citenamefont
  {Chubukov}(1973)}]{denisov1973photonuclear}%
  \BibitemOpen
  \bibfield  {author} {\bibinfo {author} {\bibfnamefont {V.~P.}\ \bibnamefont
  {Denisov}}\ and\ \bibinfo {author} {\bibfnamefont {I.~Y.}\ \bibnamefont
  {Chubukov}},\ }\href@noop {} {\bibfield  {journal} {\bibinfo  {journal}
  {Izvestiya Akademii Nauk SSSR, Seriya Fizicheskaya}\ }\textbf {\bibinfo
  {volume} {37}},\ \bibinfo {pages} {107} (\bibinfo {year} {1973})}\BibitemShut
  {NoStop}%
\bibitem [{\citenamefont {Berman}\ \emph {et~al.}(1965)\citenamefont {Berman},
  \citenamefont {Bramblett}, \citenamefont {Caldwell}, \citenamefont {Harvey},\
  and\ \citenamefont {Fultz}}]{berman1965photo}%
  \BibitemOpen
  \bibfield  {author} {\bibinfo {author} {\bibfnamefont {B.~L.}\ \bibnamefont
  {Berman}}, \bibinfo {author} {\bibfnamefont {R.~L.}\ \bibnamefont
  {Bramblett}}, \bibinfo {author} {\bibfnamefont {J.~T.}\ \bibnamefont
  {Caldwell}}, \bibinfo {author} {\bibfnamefont {R.~R.}\ \bibnamefont
  {Harvey}},\ and\ \bibinfo {author} {\bibfnamefont {S.~C.}\ \bibnamefont
  {Fultz}},\ }\href@noop {} {\bibfield  {journal} {\bibinfo  {journal}
  {Physical Review Letters}\ }\textbf {\bibinfo {volume} {15}},\ \bibinfo
  {pages} {727} (\bibinfo {year} {1965})}\BibitemShut {NoStop}%
\bibitem [{\citenamefont {Bazhanov}\ \emph {et~al.}(1965)\citenamefont
  {Bazhanov}, \citenamefont {Komar}, \citenamefont {Kulikov},\ and\
  \citenamefont {Makhnovsky}}]{bazhanov1965li6}%
  \BibitemOpen
  \bibfield  {author} {\bibinfo {author} {\bibfnamefont {E.~B.}\ \bibnamefont
  {Bazhanov}}, \bibinfo {author} {\bibfnamefont {A.~P.}\ \bibnamefont {Komar}},
  \bibinfo {author} {\bibfnamefont {A.~V.}\ \bibnamefont {Kulikov}},\ and\
  \bibinfo {author} {\bibfnamefont {E.~D.}\ \bibnamefont {Makhnovsky}},\
  }\href@noop {} {\bibfield  {journal} {\bibinfo  {journal} {Nuclear Physics}\
  }\textbf {\bibinfo {volume} {68}},\ \bibinfo {pages} {191} (\bibinfo {year}
  {1965})}\BibitemShut {NoStop}%
\bibitem [{\citenamefont {Kulchitskii}\ \emph {et~al.}(1963)\citenamefont
  {Kulchitskii}, \citenamefont {Volkov}, \citenamefont {Denisov},\ and\
  \citenamefont {Ogurtsov}}]{kulchitskii1963energy}%
  \BibitemOpen
  \bibfield  {author} {\bibinfo {author} {\bibfnamefont {L.~A.}\ \bibnamefont
  {Kulchitskii}}, \bibinfo {author} {\bibfnamefont {Y.~M.}\ \bibnamefont
  {Volkov}}, \bibinfo {author} {\bibfnamefont {V.~P.}\ \bibnamefont
  {Denisov}},\ and\ \bibinfo {author} {\bibfnamefont {V.~I.}\ \bibnamefont
  {Ogurtsov}},\ }\href@noop {} {\bibfield  {journal} {\bibinfo  {journal}
  {Rossiiskoi Akademii Nauk, Ser. Fiz}\ }\textbf {\bibinfo {volume} {27}},\
  \bibinfo {pages} {1412} (\bibinfo {year} {1963})}\BibitemShut {NoStop}%
\bibitem [{\citenamefont {Bramblett}\ \emph {et~al.}(1973)\citenamefont
  {Bramblett}, \citenamefont {Berman}, \citenamefont {Kelly}, \citenamefont
  {Caldwell},\ and\ \citenamefont {Fultz}}]{bramblett1973photoneutron}%
  \BibitemOpen
  \bibfield  {author} {\bibinfo {author} {\bibfnamefont {R.~L.}\ \bibnamefont
  {Bramblett}}, \bibinfo {author} {\bibfnamefont {B.~L.}\ \bibnamefont
  {Berman}}, \bibinfo {author} {\bibfnamefont {M.~A.}\ \bibnamefont {Kelly}},
  \bibinfo {author} {\bibfnamefont {J.~T.}\ \bibnamefont {Caldwell}},\ and\
  \bibinfo {author} {\bibfnamefont {S.~C.}\ \bibnamefont {Fultz}},\ }\href@noop
  {} {\emph {\bibinfo {title} {Photoneutron cross sections for $^{7}${L}i}}},\
  \bibinfo {type} {Tech. Rep.}\ (\bibinfo  {institution} {Lawrence Livermore
  Laboratory, University of California},\ \bibinfo {year} {1973})\BibitemShut
  {NoStop}%
\bibitem [{\citenamefont {Fuller}(1985)}]{fuller1985photonuclear}%
  \BibitemOpen
  \bibfield  {author} {\bibinfo {author} {\bibfnamefont {E.~G.}\ \bibnamefont
  {Fuller}},\ }\href@noop {} {\bibfield  {journal} {\bibinfo  {journal}
  {Physics Reports}\ }\textbf {\bibinfo {volume} {127}},\ \bibinfo {pages}
  {185} (\bibinfo {year} {1985})}\BibitemShut {NoStop}%
\bibitem [{\citenamefont {Zubanov}\ \emph {et~al.}(1992)\citenamefont
  {Zubanov}, \citenamefont {Thompson}, \citenamefont {Berman}, \citenamefont
  {Jury}, \citenamefont {Pywell},\ and\ \citenamefont {McNeill}}]{zubanov1992}%
  \BibitemOpen
  \bibfield  {author} {\bibinfo {author} {\bibfnamefont {D.}~\bibnamefont
  {Zubanov}}, \bibinfo {author} {\bibfnamefont {M.~N.}\ \bibnamefont
  {Thompson}}, \bibinfo {author} {\bibfnamefont {B.~L.}\ \bibnamefont
  {Berman}}, \bibinfo {author} {\bibfnamefont {J.~W.}\ \bibnamefont {Jury}},
  \bibinfo {author} {\bibfnamefont {R.~E.}\ \bibnamefont {Pywell}},\ and\
  \bibinfo {author} {\bibfnamefont {K.~G.}\ \bibnamefont {McNeill}},\
  }\href@noop {} {\bibfield  {journal} {\bibinfo  {journal} {Phys. Rev. C}\
  }\textbf {\bibinfo {volume} {46}},\ \bibinfo {pages} {1147} (\bibinfo {year}
  {1992})}\BibitemShut {NoStop}%
\bibitem [{\citenamefont {Jury}\ \emph {et~al.}(1980)\citenamefont {Jury},
  \citenamefont {Berman}, \citenamefont {Faul}, \citenamefont {Meyer},\ and\
  \citenamefont {Woodworth}}]{jury1980}%
  \BibitemOpen
  \bibfield  {author} {\bibinfo {author} {\bibfnamefont {J.~W.}\ \bibnamefont
  {Jury}}, \bibinfo {author} {\bibfnamefont {B.~L.}\ \bibnamefont {Berman}},
  \bibinfo {author} {\bibfnamefont {D.~D.}\ \bibnamefont {Faul}}, \bibinfo
  {author} {\bibfnamefont {P.}~\bibnamefont {Meyer}},\ and\ \bibinfo {author}
  {\bibfnamefont {J.~G.}\ \bibnamefont {Woodworth}},\ }\href@noop {} {\bibfield
   {journal} {\bibinfo  {journal} {Phys. Rev. C}\ }\textbf {\bibinfo {volume}
  {21}},\ \bibinfo {pages} {503} (\bibinfo {year} {1980})}\BibitemShut
  {NoStop}%
\bibitem [{\citenamefont {Woodworth}\ \emph {et~al.}(1979)\citenamefont
  {Woodworth}, \citenamefont {McNeill}, \citenamefont {Jury}, \citenamefont
  {Alvarez}, \citenamefont {Berman}, \citenamefont {Faul},\ and\ \citenamefont
  {Meyer}}]{woodworth1979}%
  \BibitemOpen
  \bibfield  {author} {\bibinfo {author} {\bibfnamefont {J.~G.}\ \bibnamefont
  {Woodworth}}, \bibinfo {author} {\bibfnamefont {K.~G.}\ \bibnamefont
  {McNeill}}, \bibinfo {author} {\bibfnamefont {J.~W.}\ \bibnamefont {Jury}},
  \bibinfo {author} {\bibfnamefont {R.~A.}\ \bibnamefont {Alvarez}}, \bibinfo
  {author} {\bibfnamefont {B.~L.}\ \bibnamefont {Berman}}, \bibinfo {author}
  {\bibfnamefont {D.~D.}\ \bibnamefont {Faul}},\ and\ \bibinfo {author}
  {\bibfnamefont {P.}~\bibnamefont {Meyer}},\ }\href@noop {} {\bibfield
  {journal} {\bibinfo  {journal} {Phys. Rev. C}\ }\textbf {\bibinfo {volume}
  {19}},\ \bibinfo {pages} {1667} (\bibinfo {year} {1979})}\BibitemShut
  {NoStop}%
\bibitem [{\citenamefont {Allen}\ \emph {et~al.}(1981)\citenamefont {Allen},
  \citenamefont {Muirhead},\ and\ \citenamefont {Webb}}]{allen1981}%
  \BibitemOpen
  \bibfield  {author} {\bibinfo {author} {\bibfnamefont {P.}~\bibnamefont
  {Allen}}, \bibinfo {author} {\bibfnamefont {E.}~\bibnamefont {Muirhead}},\
  and\ \bibinfo {author} {\bibfnamefont {D.}~\bibnamefont {Webb}},\ }\href@noop
  {} {\bibfield  {journal} {\bibinfo  {journal} {Nuclear Physics A}\ }\textbf
  {\bibinfo {volume} {357}},\ \bibinfo {pages} {171} (\bibinfo {year}
  {1981})}\BibitemShut {NoStop}%
\bibitem [{\citenamefont {Gorbunov}\ \emph {et~al.}(1962)\citenamefont
  {Gorbunov}, \citenamefont {Dubrovina}, \citenamefont {Osipova}, \citenamefont
  {Silaeva},\ and\ \citenamefont {Cerenkov}}]{gorbunov1962}%
  \BibitemOpen
  \bibfield  {author} {\bibinfo {author} {\bibfnamefont {A.~N.}\ \bibnamefont
  {Gorbunov}}, \bibinfo {author} {\bibfnamefont {V.~A.}\ \bibnamefont
  {Dubrovina}}, \bibinfo {author} {\bibfnamefont {V.~A.}\ \bibnamefont
  {Osipova}}, \bibinfo {author} {\bibfnamefont {V.~S.}\ \bibnamefont
  {Silaeva}},\ and\ \bibinfo {author} {\bibfnamefont {P.~A.}\ \bibnamefont
  {Cerenkov}},\ }\href@noop {} {\bibfield  {journal} {\bibinfo  {journal}
  {Journal of Experimental and Theoretical Physics}\ }\textbf {\bibinfo
  {volume} {15}},\ \bibinfo {pages} {520} (\bibinfo {year} {1962})}\BibitemShut
  {NoStop}%
\bibitem [{\citenamefont {Varlamov}\ \emph {et~al.}(1979)\citenamefont
  {Varlamov}, \citenamefont {Ishkhanov}, \citenamefont {Kapitonov},
  \citenamefont {Shvedunov},\ and\ \citenamefont
  {Prokopchuk}}]{varlamov1979effect}%
  \BibitemOpen
  \bibfield  {author} {\bibinfo {author} {\bibfnamefont {V.~V.}\ \bibnamefont
  {Varlamov}}, \bibinfo {author} {\bibfnamefont {B.~S.}\ \bibnamefont
  {Ishkhanov}}, \bibinfo {author} {\bibfnamefont {I.~M.}\ \bibnamefont
  {Kapitonov}}, \bibinfo {author} {\bibfnamefont {V.~I.}\ \bibnamefont
  {Shvedunov}},\ and\ \bibinfo {author} {\bibfnamefont {Y.~I.}\ \bibnamefont
  {Prokopchuk}},\ }\href@noop {} {\bibfield  {journal} {\bibinfo  {journal}
  {Yad. Fiz.(USSR)}\ }\textbf {\bibinfo {volume} {30}} (\bibinfo {year}
  {1979})}\BibitemShut {NoStop}%
\bibitem [{\citenamefont {Anderson}\ \emph {et~al.}(1969)\citenamefont
  {Anderson}, \citenamefont {Cook},\ and\ \citenamefont
  {Englert}}]{anderson1969}%
  \BibitemOpen
  \bibfield  {author} {\bibinfo {author} {\bibfnamefont {D.~W.}\ \bibnamefont
  {Anderson}}, \bibinfo {author} {\bibfnamefont {B.~C.}\ \bibnamefont {Cook}},\
  and\ \bibinfo {author} {\bibfnamefont {T.~J.}\ \bibnamefont {Englert}},\
  }\href@noop {} {\bibfield  {journal} {\bibinfo  {journal} {Nuclear Physics
  A}\ }\textbf {\bibinfo {volume} {127}},\ \bibinfo {pages} {474} (\bibinfo
  {year} {1969})}\BibitemShut {NoStop}%
\bibitem [{\citenamefont {Miorelli}\ \emph {et~al.}(2016)\citenamefont
  {Miorelli}, \citenamefont {Bacca}, \citenamefont {Barnea}, \citenamefont
  {Hagen}, \citenamefont {Jansen}, \citenamefont {Orlandini},\ and\
  \citenamefont {Papenbrock}}]{miorelli2016electric}%
  \BibitemOpen
  \bibfield  {author} {\bibinfo {author} {\bibfnamefont {M.}~\bibnamefont
  {Miorelli}}, \bibinfo {author} {\bibfnamefont {S.}~\bibnamefont {Bacca}},
  \bibinfo {author} {\bibfnamefont {N.}~\bibnamefont {Barnea}}, \bibinfo
  {author} {\bibfnamefont {G.}~\bibnamefont {Hagen}}, \bibinfo {author}
  {\bibfnamefont {G.~R.}\ \bibnamefont {Jansen}}, \bibinfo {author}
  {\bibfnamefont {G.}~\bibnamefont {Orlandini}},\ and\ \bibinfo {author}
  {\bibfnamefont {T.}~\bibnamefont {Papenbrock}},\ }\href@noop {} {\bibfield
  {journal} {\bibinfo  {journal} {Physical Review C}\ }\textbf {\bibinfo
  {volume} {94}},\ \bibinfo {pages} {034317} (\bibinfo {year}
  {2016})}\BibitemShut {NoStop}%
\bibitem [{\citenamefont {Orce}\ \emph {et~al.}(2012)\citenamefont {Orce},
  \citenamefont {Drake}, \citenamefont {Djongolov}, \citenamefont
  {Navr\'atil},\ and\ \citenamefont {{\it et al.}}}]{orce2012reorientation}%
  \BibitemOpen
  \bibfield  {author} {\bibinfo {author} {\bibfnamefont {J.~N.}\ \bibnamefont
  {Orce}}, \bibinfo {author} {\bibfnamefont {T.~E.}\ \bibnamefont {Drake}},
  \bibinfo {author} {\bibfnamefont {M.~K.}\ \bibnamefont {Djongolov}}, \bibinfo
  {author} {\bibfnamefont {P.}~\bibnamefont {Navr\'atil}},\ and\ \bibinfo
  {author} {\bibnamefont {{\it et al.}}},\ }\href@noop {} {\bibfield  {journal}
  {\bibinfo  {journal} {Physical Review C}\ }\textbf {\bibinfo {volume} {86}},\
  \bibinfo {pages} {041303} (\bibinfo {year} {2012})}\BibitemShut {NoStop}%
\bibitem [{\citenamefont {Kumar-Raju}\ \emph {et~al.}(2018)\citenamefont
  {Kumar-Raju}, \citenamefont {Orce}, \citenamefont {Navr{\'a}til},
  \citenamefont {Ball}, \citenamefont {Drake}, \citenamefont {Triambak},
  \citenamefont {Hackman}, \citenamefont {Pearson}, \citenamefont {Abrahams},
  \citenamefont {Akakpo} \emph {et~al.}}]{raju2018reorientation}%
  \BibitemOpen
  \bibfield  {author} {\bibinfo {author} {\bibfnamefont {M.}~\bibnamefont
  {Kumar-Raju}}, \bibinfo {author} {\bibfnamefont {J.~N.}\ \bibnamefont
  {Orce}}, \bibinfo {author} {\bibfnamefont {P.}~\bibnamefont {Navr{\'a}til}},
  \bibinfo {author} {\bibfnamefont {G.~C.}\ \bibnamefont {Ball}}, \bibinfo
  {author} {\bibfnamefont {T.~E.}\ \bibnamefont {Drake}}, \bibinfo {author}
  {\bibfnamefont {S.}~\bibnamefont {Triambak}}, \bibinfo {author}
  {\bibfnamefont {G.}~\bibnamefont {Hackman}}, \bibinfo {author} {\bibfnamefont
  {C.~J.}\ \bibnamefont {Pearson}}, \bibinfo {author} {\bibfnamefont {K.~J.}\
  \bibnamefont {Abrahams}}, \bibinfo {author} {\bibfnamefont {E.~H.}\
  \bibnamefont {Akakpo}}, \emph {et~al.},\ }\href@noop {} {\bibfield  {journal}
  {\bibinfo  {journal} {Physics Letters B}\ }\textbf {\bibinfo {volume}
  {777}},\ \bibinfo {pages} {250} (\bibinfo {year} {2018})}\BibitemShut
  {NoStop}%
\bibitem [{\citenamefont {M{\"o}ller}\ \emph {et~al.}(1995)\citenamefont
  {M{\"o}ller}, \citenamefont {Nix} \emph {et~al.}}]{moller1995atomic}%
  \BibitemOpen
  \bibfield  {author} {\bibinfo {author} {\bibfnamefont {P.}~\bibnamefont
  {M{\"o}ller}}, \bibinfo {author} {\bibfnamefont {J.~R.}\ \bibnamefont {Nix}},
  \emph {et~al.},\ }\href@noop {} {\bibfield  {journal} {\bibinfo  {journal}
  {Atomic Data Nuclear Data Tables}\ }\textbf {\bibinfo {volume} {66}},\
  \bibinfo {pages} {131} (\bibinfo {year} {1995})}\BibitemShut {NoStop}%
\bibitem [{\citenamefont {Gibelin}\ \emph {et~al.}(2008)\citenamefont
  {Gibelin}, \citenamefont {Beaumel}, \citenamefont {Motobayashi},
  \citenamefont {Blumenfeld}, \citenamefont {Aoi}, \citenamefont {Baba},
  \citenamefont {Elekes}, \citenamefont {Fortier}, \citenamefont {Frascaria},
  \citenamefont {Fukuda} \emph {et~al.}}]{gibelin2008decay}%
  \BibitemOpen
  \bibfield  {author} {\bibinfo {author} {\bibfnamefont {J.}~\bibnamefont
  {Gibelin}}, \bibinfo {author} {\bibfnamefont {D.}~\bibnamefont {Beaumel}},
  \bibinfo {author} {\bibfnamefont {T.}~\bibnamefont {Motobayashi}}, \bibinfo
  {author} {\bibfnamefont {Y.}~\bibnamefont {Blumenfeld}}, \bibinfo {author}
  {\bibfnamefont {N.}~\bibnamefont {Aoi}}, \bibinfo {author} {\bibfnamefont
  {H.}~\bibnamefont {Baba}}, \bibinfo {author} {\bibfnamefont {Z.}~\bibnamefont
  {Elekes}}, \bibinfo {author} {\bibfnamefont {S.}~\bibnamefont {Fortier}},
  \bibinfo {author} {\bibfnamefont {N.}~\bibnamefont {Frascaria}}, \bibinfo
  {author} {\bibfnamefont {N.}~\bibnamefont {Fukuda}}, \emph {et~al.},\
  }\href@noop {} {\bibfield  {journal} {\bibinfo  {journal} {Physical review
  letters}\ }\textbf {\bibinfo {volume} {101}},\ \bibinfo {pages} {212503}
  (\bibinfo {year} {2008})}\BibitemShut {NoStop}%
\bibitem [{\citenamefont {Paar}\ \emph {et~al.}(2007)\citenamefont {Paar},
  \citenamefont {Vretenar}, \citenamefont {Khan},\ and\ \citenamefont
  {Colo}}]{paar2007exotic}%
  \BibitemOpen
  \bibfield  {author} {\bibinfo {author} {\bibfnamefont {N.}~\bibnamefont
  {Paar}}, \bibinfo {author} {\bibfnamefont {D.}~\bibnamefont {Vretenar}},
  \bibinfo {author} {\bibfnamefont {E.}~\bibnamefont {Khan}},\ and\ \bibinfo
  {author} {\bibfnamefont {G.}~\bibnamefont {Colo}},\ }\href@noop {} {\bibfield
   {journal} {\bibinfo  {journal} {Reports on Progress in Physics}\ }\textbf
  {\bibinfo {volume} {70}},\ \bibinfo {pages} {691} (\bibinfo {year}
  {2007})}\BibitemShut {NoStop}%
\bibitem [{\citenamefont {von Neumann-Cosel}(2016)}]{von2016comment}%
  \BibitemOpen
  \bibfield  {author} {\bibinfo {author} {\bibfnamefont {P.}~\bibnamefont {von
  Neumann-Cosel}},\ }\href@noop {} {\bibfield  {journal} {\bibinfo  {journal}
  {Physical Review C}\ }\textbf {\bibinfo {volume} {93}},\ \bibinfo {pages}
  {049801} (\bibinfo {year} {2016})}\BibitemShut {NoStop}%
\bibitem [{\citenamefont {Arsenyev}\ \emph {et~al.}(2016)\citenamefont
  {Arsenyev}, \citenamefont {Severyukhin}, \citenamefont {Voronov},\ and\
  \citenamefont {Van~Giai}}]{arsenyev2016effects}%
  \BibitemOpen
  \bibfield  {author} {\bibinfo {author} {\bibfnamefont {N.}~\bibnamefont
  {Arsenyev}}, \bibinfo {author} {\bibfnamefont {A.}~\bibnamefont
  {Severyukhin}}, \bibinfo {author} {\bibfnamefont {V.}~\bibnamefont
  {Voronov}},\ and\ \bibinfo {author} {\bibfnamefont {N.}~\bibnamefont
  {Van~Giai}},\ }in\ \href@noop {} {\emph {\bibinfo {booktitle} {EPJ Web of
  Conferences}}},\ Vol.\ \bibinfo {volume} {107}\ (\bibinfo {organization} {EDP
  Sciences},\ \bibinfo {year} {2016})\ p.\ \bibinfo {pages} {05006}\BibitemShut
  {NoStop}%
\bibitem [{\citenamefont {Cook}(1957)}]{cook1957photodisintegration}%
  \BibitemOpen
  \bibfield  {author} {\bibinfo {author} {\bibfnamefont {B.~C.}\ \bibnamefont
  {Cook}},\ }\href@noop {} {\bibfield  {journal} {\bibinfo  {journal} {Physical
  Review}\ }\textbf {\bibinfo {volume} {106}},\ \bibinfo {pages} {300}
  (\bibinfo {year} {1957})}\BibitemShut {NoStop}%
\bibitem [{\citenamefont {Eramzhyan}\ \emph {et~al.}(1986)\citenamefont
  {Eramzhyan}, \citenamefont {Ishkhanov}, \citenamefont {Kapitonov},\ and\
  \citenamefont {Neudatchin}}]{eramzhyan1986giant}%
  \BibitemOpen
  \bibfield  {author} {\bibinfo {author} {\bibfnamefont {R.~A.}\ \bibnamefont
  {Eramzhyan}}, \bibinfo {author} {\bibfnamefont {B.~S.}\ \bibnamefont
  {Ishkhanov}}, \bibinfo {author} {\bibfnamefont {I.~M.}\ \bibnamefont
  {Kapitonov}},\ and\ \bibinfo {author} {\bibfnamefont {V.~G.}\ \bibnamefont
  {Neudatchin}},\ }\href@noop {} {\bibfield  {journal} {\bibinfo  {journal}
  {Physics Reports}\ }\textbf {\bibinfo {volume} {136}},\ \bibinfo {pages}
  {229} (\bibinfo {year} {1986})}\BibitemShut {NoStop}%
\bibitem [{\citenamefont {Nakayama}\ \emph {et~al.}(2001)\citenamefont
  {Nakayama}, \citenamefont {Yamagata}, \citenamefont {Akimune}, \citenamefont
  {Daito}, \citenamefont {Fujimura}, \citenamefont {Fujita}, \citenamefont
  {Fujiwara}, \citenamefont {Fushimi}, \citenamefont {Greenfield},
  \citenamefont {Kohri} \emph {et~al.}}]{nakayama2001dipole}%
  \BibitemOpen
  \bibfield  {author} {\bibinfo {author} {\bibfnamefont {S.}~\bibnamefont
  {Nakayama}}, \bibinfo {author} {\bibfnamefont {T.}~\bibnamefont {Yamagata}},
  \bibinfo {author} {\bibfnamefont {H.}~\bibnamefont {Akimune}}, \bibinfo
  {author} {\bibfnamefont {I.}~\bibnamefont {Daito}}, \bibinfo {author}
  {\bibfnamefont {H.}~\bibnamefont {Fujimura}}, \bibinfo {author}
  {\bibfnamefont {Y.}~\bibnamefont {Fujita}}, \bibinfo {author} {\bibfnamefont
  {M.}~\bibnamefont {Fujiwara}}, \bibinfo {author} {\bibfnamefont
  {K.}~\bibnamefont {Fushimi}}, \bibinfo {author} {\bibfnamefont
  {M.}~\bibnamefont {Greenfield}}, \bibinfo {author} {\bibfnamefont
  {H.}~\bibnamefont {Kohri}}, \emph {et~al.},\ }\href@noop {} {\bibfield
  {journal} {\bibinfo  {journal} {Physical Review Letters}\ }\textbf {\bibinfo
  {volume} {87}},\ \bibinfo {pages} {122502} (\bibinfo {year}
  {2001})}\BibitemShut {NoStop}%
\bibitem [{\citenamefont {Burda}\ \emph {et~al.}(2010)\citenamefont {Burda},
  \citenamefont {von Neumann-Cosel}, \citenamefont {Richter}, \citenamefont
  {Forss{\'e}n},\ and\ \citenamefont {Brown}}]{burda2010resonance}%
  \BibitemOpen
  \bibfield  {author} {\bibinfo {author} {\bibfnamefont {O.}~\bibnamefont
  {Burda}}, \bibinfo {author} {\bibfnamefont {P.}~\bibnamefont {von
  Neumann-Cosel}}, \bibinfo {author} {\bibfnamefont {A.}~\bibnamefont
  {Richter}}, \bibinfo {author} {\bibfnamefont {C.}~\bibnamefont
  {Forss{\'e}n}},\ and\ \bibinfo {author} {\bibfnamefont {B.~A.}\ \bibnamefont
  {Brown}},\ }\href@noop {} {\bibfield  {journal} {\bibinfo  {journal}
  {Physical Review C}\ }\textbf {\bibinfo {volume} {82}},\ \bibinfo {pages}
  {015808} (\bibinfo {year} {2010})}\BibitemShut {NoStop}%
\bibitem [{\citenamefont {Aumann}\ \emph {et~al.}(1999)\citenamefont {Aumann},
  \citenamefont {Leistenschneider}, \citenamefont {Boretzky}, \citenamefont
  {Cortina}, \citenamefont {Cub}, \citenamefont {Dostal}, \citenamefont
  {Eberlein}, \citenamefont {Elze}, \citenamefont {Emling}, \citenamefont
  {Geissel} \emph {et~al.}}]{aumann1999giant}%
  \BibitemOpen
  \bibfield  {author} {\bibinfo {author} {\bibfnamefont {T.}~\bibnamefont
  {Aumann}}, \bibinfo {author} {\bibfnamefont {A.}~\bibnamefont
  {Leistenschneider}}, \bibinfo {author} {\bibfnamefont {K.}~\bibnamefont
  {Boretzky}}, \bibinfo {author} {\bibfnamefont {D.}~\bibnamefont {Cortina}},
  \bibinfo {author} {\bibfnamefont {J.}~\bibnamefont {Cub}}, \bibinfo {author}
  {\bibfnamefont {W.}~\bibnamefont {Dostal}}, \bibinfo {author} {\bibfnamefont
  {B.}~\bibnamefont {Eberlein}}, \bibinfo {author} {\bibfnamefont {T.~W.}\
  \bibnamefont {Elze}}, \bibinfo {author} {\bibfnamefont {H.}~\bibnamefont
  {Emling}}, \bibinfo {author} {\bibfnamefont {H.}~\bibnamefont {Geissel}},
  \emph {et~al.},\ }\href@noop {} {\bibfield  {journal} {\bibinfo  {journal}
  {Nuclear Physics A}\ }\textbf {\bibinfo {volume} {649}},\ \bibinfo {pages}
  {297} (\bibinfo {year} {1999})}\BibitemShut {NoStop}%
\bibitem [{\citenamefont {Leistenschneider}\ \emph {et~al.}(2001)\citenamefont
  {Leistenschneider}, \citenamefont {Aumann}, \citenamefont {Boretzky},
  \citenamefont {Cortina}, \citenamefont {Cub}, \citenamefont {Pramanik},
  \citenamefont {Dostal}, \citenamefont {Elze}, \citenamefont {Emling},
  \citenamefont {Geissel} \emph {et~al.}}]{leistenschneider2001photoneutron}%
  \BibitemOpen
  \bibfield  {author} {\bibinfo {author} {\bibfnamefont {A.}~\bibnamefont
  {Leistenschneider}}, \bibinfo {author} {\bibfnamefont {T.}~\bibnamefont
  {Aumann}}, \bibinfo {author} {\bibfnamefont {K.}~\bibnamefont {Boretzky}},
  \bibinfo {author} {\bibfnamefont {D.}~\bibnamefont {Cortina}}, \bibinfo
  {author} {\bibfnamefont {J.}~\bibnamefont {Cub}}, \bibinfo {author}
  {\bibfnamefont {U.~D.}\ \bibnamefont {Pramanik}}, \bibinfo {author}
  {\bibfnamefont {W.}~\bibnamefont {Dostal}}, \bibinfo {author} {\bibfnamefont
  {T.~W.}\ \bibnamefont {Elze}}, \bibinfo {author} {\bibfnamefont
  {H.}~\bibnamefont {Emling}}, \bibinfo {author} {\bibfnamefont
  {H.}~\bibnamefont {Geissel}}, \emph {et~al.},\ }\href@noop {} {\bibfield
  {journal} {\bibinfo  {journal} {Physical review letters}\ }\textbf {\bibinfo
  {volume} {86}},\ \bibinfo {pages} {5442} (\bibinfo {year}
  {2001})}\BibitemShut {NoStop}%
\bibitem [{\citenamefont {Terasaki}\ and\ \citenamefont
  {Engel}(2006)}]{terasaki2006self}%
  \BibitemOpen
  \bibfield  {author} {\bibinfo {author} {\bibfnamefont {J.}~\bibnamefont
  {Terasaki}}\ and\ \bibinfo {author} {\bibfnamefont {J.}~\bibnamefont
  {Engel}},\ }\href@noop {} {\bibfield  {journal} {\bibinfo  {journal}
  {Physical Review C}\ }\textbf {\bibinfo {volume} {74}},\ \bibinfo {pages}
  {044301} (\bibinfo {year} {2006})}\BibitemShut {NoStop}%
\bibitem [{\citenamefont {B{\"u}rger}\ \emph {et~al.}(2012)\citenamefont
  {B{\"u}rger}, \citenamefont {Larsen}, \citenamefont {Hilaire}, \citenamefont
  {Guttormsen}, \citenamefont {Harissopulos}, \citenamefont {Kmiecik},
  \citenamefont {Konstantinopoulos}, \citenamefont {Krti{\v{c}}ka},
  \citenamefont {Lagoyannis}, \citenamefont {L{\"o}nnroth} \emph
  {et~al.}}]{burger2012nuclear}%
  \BibitemOpen
  \bibfield  {author} {\bibinfo {author} {\bibfnamefont {A.}~\bibnamefont
  {B{\"u}rger}}, \bibinfo {author} {\bibfnamefont {A.}~\bibnamefont {Larsen}},
  \bibinfo {author} {\bibfnamefont {S.}~\bibnamefont {Hilaire}}, \bibinfo
  {author} {\bibfnamefont {M.}~\bibnamefont {Guttormsen}}, \bibinfo {author}
  {\bibfnamefont {S.}~\bibnamefont {Harissopulos}}, \bibinfo {author}
  {\bibfnamefont {M.}~\bibnamefont {Kmiecik}}, \bibinfo {author} {\bibfnamefont
  {T.}~\bibnamefont {Konstantinopoulos}}, \bibinfo {author} {\bibfnamefont
  {M.}~\bibnamefont {Krti{\v{c}}ka}}, \bibinfo {author} {\bibfnamefont
  {A.}~\bibnamefont {Lagoyannis}}, \bibinfo {author} {\bibfnamefont
  {T.}~\bibnamefont {L{\"o}nnroth}}, \emph {et~al.},\ }\href@noop {} {\bibfield
   {journal} {\bibinfo  {journal} {Physical Review C}\ }\textbf {\bibinfo
  {volume} {85}},\ \bibinfo {pages} {064328} (\bibinfo {year}
  {2012})}\BibitemShut {NoStop}%
\bibitem [{\citenamefont {Larsen}\ \emph {et~al.}(2007)\citenamefont {Larsen},
  \citenamefont {Guttormsen}, \citenamefont {Chankova}, \citenamefont
  {Ingebretsen}, \citenamefont {L{\"o}nnroth}, \citenamefont {Messelt},
  \citenamefont {Rekstad}, \citenamefont {Schiller}, \citenamefont {Siem},
  \citenamefont {Syed} \emph {et~al.}}]{larsen2007nuclear}%
  \BibitemOpen
  \bibfield  {author} {\bibinfo {author} {\bibfnamefont {A.~C.}\ \bibnamefont
  {Larsen}}, \bibinfo {author} {\bibfnamefont {M.}~\bibnamefont {Guttormsen}},
  \bibinfo {author} {\bibfnamefont {R.}~\bibnamefont {Chankova}}, \bibinfo
  {author} {\bibfnamefont {F.}~\bibnamefont {Ingebretsen}}, \bibinfo {author}
  {\bibfnamefont {T.}~\bibnamefont {L{\"o}nnroth}}, \bibinfo {author}
  {\bibfnamefont {S.}~\bibnamefont {Messelt}}, \bibinfo {author} {\bibfnamefont
  {J.}~\bibnamefont {Rekstad}}, \bibinfo {author} {\bibfnamefont
  {A.}~\bibnamefont {Schiller}}, \bibinfo {author} {\bibfnamefont
  {S.}~\bibnamefont {Siem}}, \bibinfo {author} {\bibfnamefont {N.~U.~H.}\
  \bibnamefont {Syed}}, \emph {et~al.},\ }\href@noop {} {\bibfield  {journal}
  {\bibinfo  {journal} {Physical Review C}\ }\textbf {\bibinfo {volume} {76}},\
  \bibinfo {pages} {044303} (\bibinfo {year} {2007})}\BibitemShut {NoStop}%
\bibitem [{\citenamefont {Larsen}\ \emph {et~al.}(2019)\citenamefont {Larsen},
  \citenamefont {Spyrou}, \citenamefont {Liddick},\ and\ \citenamefont
  {Guttormsen}}]{larsen2019novel}%
  \BibitemOpen
  \bibfield  {author} {\bibinfo {author} {\bibfnamefont {A.-C.}\ \bibnamefont
  {Larsen}}, \bibinfo {author} {\bibfnamefont {A.}~\bibnamefont {Spyrou}},
  \bibinfo {author} {\bibfnamefont {S.~N.}\ \bibnamefont {Liddick}},\ and\
  \bibinfo {author} {\bibfnamefont {M.}~\bibnamefont {Guttormsen}},\
  }\href@noop {} {\bibfield  {journal} {\bibinfo  {journal} {Progress in
  Particle and Nuclear Physics}\ }\textbf {\bibinfo {volume} {107}},\ \bibinfo
  {pages} {69} (\bibinfo {year} {2019})}\BibitemShut {NoStop}%
\bibitem [{\citenamefont {Midtb{\o}}\ \emph {et~al.}(2021)\citenamefont
  {Midtb{\o}}, \citenamefont {Zeiser}, \citenamefont {Lima}, \citenamefont
  {Larsen}, \citenamefont {Tveten}, \citenamefont {Guttormsen}, \citenamefont
  {Garrote}, \citenamefont {Kvellestad},\ and\ \citenamefont
  {Renstr{\o}m}}]{midtbo2021new}%
  \BibitemOpen
  \bibfield  {author} {\bibinfo {author} {\bibfnamefont {J.~E.}\ \bibnamefont
  {Midtb{\o}}}, \bibinfo {author} {\bibfnamefont {F.}~\bibnamefont {Zeiser}},
  \bibinfo {author} {\bibfnamefont {E.}~\bibnamefont {Lima}}, \bibinfo {author}
  {\bibfnamefont {A.-C.}\ \bibnamefont {Larsen}}, \bibinfo {author}
  {\bibfnamefont {G.~M.}\ \bibnamefont {Tveten}}, \bibinfo {author}
  {\bibfnamefont {M.}~\bibnamefont {Guttormsen}}, \bibinfo {author}
  {\bibfnamefont {F.~L.~B.}\ \bibnamefont {Garrote}}, \bibinfo {author}
  {\bibfnamefont {A.}~\bibnamefont {Kvellestad}},\ and\ \bibinfo {author}
  {\bibfnamefont {T.}~\bibnamefont {Renstr{\o}m}},\ }\href@noop {} {\bibfield
  {journal} {\bibinfo  {journal} {Computer Physics Communications}\ }\textbf
  {\bibinfo {volume} {262}},\ \bibinfo {pages} {107795} (\bibinfo {year}
  {2021})}\BibitemShut {NoStop}%
\bibitem [{\citenamefont {Zilges}\ \emph {et~al.}(2022)\citenamefont {Zilges},
  \citenamefont {Balabanski}, \citenamefont {Isaak},\ and\ \citenamefont
  {Pietralla}}]{zilges2022photonuclear}%
  \BibitemOpen
  \bibfield  {author} {\bibinfo {author} {\bibfnamefont {A.}~\bibnamefont
  {Zilges}}, \bibinfo {author} {\bibfnamefont {D.~L.}\ \bibnamefont
  {Balabanski}}, \bibinfo {author} {\bibfnamefont {J.}~\bibnamefont {Isaak}},\
  and\ \bibinfo {author} {\bibfnamefont {N.}~\bibnamefont {Pietralla}},\
  }\href@noop {} {\bibfield  {journal} {\bibinfo  {journal} {Progress in
  Particle and Nuclear Physics}\ }\textbf {\bibinfo {volume} {122}},\ \bibinfo
  {pages} {103903} (\bibinfo {year} {2022})}\BibitemShut {NoStop}%
\bibitem [{\citenamefont {Ngwetsheni}\ and\ \citenamefont
  {Orce}(2019{\natexlab{a}})}]{ngwetsheni2019continuing}%
  \BibitemOpen
  \bibfield  {author} {\bibinfo {author} {\bibfnamefont {C.}~\bibnamefont
  {Ngwetsheni}}\ and\ \bibinfo {author} {\bibfnamefont {J.~N.}\ \bibnamefont
  {Orce}},\ }\href@noop {} {\bibfield  {journal} {\bibinfo  {journal} {Physics
  Letters B}\ }\textbf {\bibinfo {volume} {792}},\ \bibinfo {pages} {335}
  (\bibinfo {year} {2019}{\natexlab{a}})}\BibitemShut {NoStop}%
\bibitem [{\citenamefont {Ngwetsheni}\ and\ \citenamefont
  {Orce}(2019{\natexlab{b}})}]{ngwetsheni2019combined}%
  \BibitemOpen
  \bibfield  {author} {\bibinfo {author} {\bibfnamefont {C.}~\bibnamefont
  {Ngwetsheni}}\ and\ \bibinfo {author} {\bibfnamefont {J.~N.}\ \bibnamefont
  {Orce}},\ }\href@noop {} {\bibfield  {journal} {\bibinfo  {journal}
  {Hyperfine Interactions}\ }\textbf {\bibinfo {volume} {240}},\ \bibinfo
  {pages} {94} (\bibinfo {year} {2019}{\natexlab{b}})}\BibitemShut {NoStop}%
\bibitem [{\citenamefont {Ngwetsheni}\ and\ \citenamefont
  {Orce}(2019{\natexlab{c}})}]{ngwetsheni2019how}%
  \BibitemOpen
  \bibfield  {author} {\bibinfo {author} {\bibfnamefont {C.}~\bibnamefont
  {Ngwetsheni}}\ and\ \bibinfo {author} {\bibfnamefont {J.~N.}\ \bibnamefont
  {Orce}},\ }\href@noop {} {\bibfield  {journal} {\bibinfo  {journal} {EPJ Web
  Conf.}\ }\textbf {\bibinfo {volume} {223}},\ \bibinfo {pages} {01045}
  (\bibinfo {year} {2019}{\natexlab{c}})}\BibitemShut {NoStop}%
\bibitem [{\citenamefont {Orce}(2015)}]{orce2015new}%
  \BibitemOpen
  \bibfield  {author} {\bibinfo {author} {\bibfnamefont {J.~N.}\ \bibnamefont
  {Orce}},\ }\href@noop {} {\bibfield  {journal} {\bibinfo  {journal} {Physical
  Review C}\ }\textbf {\bibinfo {volume} {91}},\ \bibinfo {pages} {064602}
  (\bibinfo {year} {2015})}\BibitemShut {NoStop}%
\bibitem [{\citenamefont {Orce}(2016)}]{orce2016reply}%
  \BibitemOpen
  \bibfield  {author} {\bibinfo {author} {\bibfnamefont {J.~N.}\ \bibnamefont
  {Orce}},\ }\href@noop {} {\bibfield  {journal} {\bibinfo  {journal} {Physical
  Review C}\ }\textbf {\bibinfo {volume} {93}},\ \bibinfo {pages} {049802}
  (\bibinfo {year} {2016})}\BibitemShut {NoStop}%
\bibitem [{exf()}]{exfor}%
  \BibitemOpen
  \href@noop {} {\bibinfo {title} {{EXFOR}: Experimental nuclear reaction
  data}},\ \bibinfo {howpublished}
  {\url{https://www-nds.iaea.org/exfor/exfor.htm}},\ \bibinfo {note} {accessed:
  2022-05-21}\BibitemShut {NoStop}%
\bibitem [{END()}]{ENDF}%
  \BibitemOpen
  \href@noop {} {\bibinfo {title} {{ENDF}: Evaluated nuclear data file}},\
  \bibinfo {howpublished} {\url{https://www-nds.iaea.org/exfor/endf.htm}},\
  \bibinfo {note} {accessed: 2022-05-21}\BibitemShut {NoStop}%
\bibitem [{\citenamefont {Goryachev}\ \emph
  {et~al.}(1968{\natexlab{a}})\citenamefont {Goryachev}, \citenamefont
  {Ishkhanov}, \citenamefont {Kapitonov}, \citenamefont {Piskarev},
  \citenamefont {Shevchen},\ and\ \citenamefont
  {Shevchen}}]{goryachev1968cross}%
  \BibitemOpen
  \bibfield  {author} {\bibinfo {author} {\bibfnamefont {B.~I.}\ \bibnamefont
  {Goryachev}}, \bibinfo {author} {\bibfnamefont {B.~S.}\ \bibnamefont
  {Ishkhanov}}, \bibinfo {author} {\bibfnamefont {I.~M.}\ \bibnamefont
  {Kapitonov}}, \bibinfo {author} {\bibfnamefont {I.~M.}\ \bibnamefont
  {Piskarev}}, \bibinfo {author} {\bibfnamefont {O.~P.}\ \bibnamefont
  {Shevchen}},\ and\ \bibinfo {author} {\bibfnamefont {V.~G.}\ \bibnamefont
  {Shevchen}},\ }\href@noop {} {\bibfield  {journal} {\bibinfo  {journal}
  {Soviet Journal of Nuclear Physics-USSR}\ }\textbf {\bibinfo {volume} {7}},\
  \bibinfo {pages} {567} (\bibinfo {year} {1968}{\natexlab{a}})}\BibitemShut
  {NoStop}%
\bibitem [{\citenamefont {Goryachev}\ \emph
  {et~al.}(1968{\natexlab{b}})\citenamefont {Goryachev}, \citenamefont
  {Ishkhanov}, \citenamefont {Shevchenko},\ and\ \citenamefont
  {Yurev}}]{goryachev1968structure}%
  \BibitemOpen
  \bibfield  {author} {\bibinfo {author} {\bibfnamefont {B.~I.}\ \bibnamefont
  {Goryachev}}, \bibinfo {author} {\bibfnamefont {B.~S.}\ \bibnamefont
  {Ishkhanov}}, \bibinfo {author} {\bibfnamefont {V.~G.}\ \bibnamefont
  {Shevchenko}},\ and\ \bibinfo {author} {\bibfnamefont {B.~A.}\ \bibnamefont
  {Yurev}},\ }\href@noop {} {\bibfield  {journal} {\bibinfo  {journal} {Soviet
  Journal of Nuclear Physics-USSR}\ }\textbf {\bibinfo {volume} {7}},\ \bibinfo
  {pages} {698} (\bibinfo {year} {1968}{\natexlab{b}})}\BibitemShut {NoStop}%
\bibitem [{\citenamefont {Navratil}(2007)}]{navratil2007local}%
  \BibitemOpen
  \bibfield  {author} {\bibinfo {author} {\bibfnamefont {P.}~\bibnamefont
  {Navratil}},\ }\href@noop {} {\bibfield  {journal} {\bibinfo  {journal}
  {Few-Body Systems}\ }\textbf {\bibinfo {volume} {41}},\ \bibinfo {pages}
  {117} (\bibinfo {year} {2007})}\BibitemShut {NoStop}%
\bibitem [{\citenamefont {Roth}\ \emph {et~al.}(2014)\citenamefont {Roth},
  \citenamefont {Calci}, \citenamefont {Langhammer},\ and\ \citenamefont
  {Binder}}]{roth2014evolved}%
  \BibitemOpen
  \bibfield  {author} {\bibinfo {author} {\bibfnamefont {R.}~\bibnamefont
  {Roth}}, \bibinfo {author} {\bibfnamefont {A.}~\bibnamefont {Calci}},
  \bibinfo {author} {\bibfnamefont {J.}~\bibnamefont {Langhammer}},\ and\
  \bibinfo {author} {\bibfnamefont {S.}~\bibnamefont {Binder}},\ }\href@noop {}
  {\bibfield  {journal} {\bibinfo  {journal} {Physical Review C}\ }\textbf
  {\bibinfo {volume} {90}},\ \bibinfo {pages} {024325} (\bibinfo {year}
  {2014})}\BibitemShut {NoStop}%
\bibitem [{\citenamefont {Entem}\ and\ \citenamefont
  {Machleidt}(2003)}]{entem2003accurate}%
  \BibitemOpen
  \bibfield  {author} {\bibinfo {author} {\bibfnamefont {D.}~\bibnamefont
  {Entem}}\ and\ \bibinfo {author} {\bibfnamefont {R.}~\bibnamefont
  {Machleidt}},\ }\href@noop {} {\bibfield  {journal} {\bibinfo  {journal}
  {Physical Review C}\ }\textbf {\bibinfo {volume} {68}},\ \bibinfo {pages}
  {041001} (\bibinfo {year} {2003})}\BibitemShut {NoStop}%
\bibitem [{\citenamefont {Entem}\ \emph {et~al.}(2017)\citenamefont {Entem},
  \citenamefont {Machleidt},\ and\ \citenamefont {Nosyk}}]{entem2017high}%
  \BibitemOpen
  \bibfield  {author} {\bibinfo {author} {\bibfnamefont {D.}~\bibnamefont
  {Entem}}, \bibinfo {author} {\bibfnamefont {R.}~\bibnamefont {Machleidt}},\
  and\ \bibinfo {author} {\bibfnamefont {Y.}~\bibnamefont {Nosyk}},\
  }\href@noop {} {\bibfield  {journal} {\bibinfo  {journal} {Physical Review
  C}\ }\textbf {\bibinfo {volume} {96}},\ \bibinfo {pages} {024004} (\bibinfo
  {year} {2017})}\BibitemShut {NoStop}%
\bibitem [{\citenamefont {Entem}\ \emph {et~al.}(2015)\citenamefont {Entem},
  \citenamefont {Kaiser}, \citenamefont {Machleidt},\ and\ \citenamefont
  {Nosyk}}]{entem2015peripheral}%
  \BibitemOpen
  \bibfield  {author} {\bibinfo {author} {\bibfnamefont {D.}~\bibnamefont
  {Entem}}, \bibinfo {author} {\bibfnamefont {N.}~\bibnamefont {Kaiser}},
  \bibinfo {author} {\bibfnamefont {R.}~\bibnamefont {Machleidt}},\ and\
  \bibinfo {author} {\bibfnamefont {Y.}~\bibnamefont {Nosyk}},\ }\href@noop {}
  {\bibfield  {journal} {\bibinfo  {journal} {Physical Review C}\ }\textbf
  {\bibinfo {volume} {91}},\ \bibinfo {pages} {014002} (\bibinfo {year}
  {2015})}\BibitemShut {NoStop}%
\bibitem [{\citenamefont {Bogner}\ \emph {et~al.}(2007)\citenamefont {Bogner},
  \citenamefont {Furnstahl},\ and\ \citenamefont
  {Perry}}]{bogner2007similarity}%
  \BibitemOpen
  \bibfield  {author} {\bibinfo {author} {\bibfnamefont {S.}~\bibnamefont
  {Bogner}}, \bibinfo {author} {\bibfnamefont {R.~J.}\ \bibnamefont
  {Furnstahl}},\ and\ \bibinfo {author} {\bibfnamefont {R.}~\bibnamefont
  {Perry}},\ }\href@noop {} {\bibfield  {journal} {\bibinfo  {journal}
  {Physical Review C}\ }\textbf {\bibinfo {volume} {75}},\ \bibinfo {pages}
  {061001} (\bibinfo {year} {2007})}\BibitemShut {NoStop}%
\bibitem [{\citenamefont {Stetcu}\ \emph {et~al.}(2009)\citenamefont {Stetcu},
  \citenamefont {Quaglioni}, \citenamefont {Friar}, \citenamefont {Hayes},\
  and\ \citenamefont {Navr{\'a}til}}]{stetcu2009electric}%
  \BibitemOpen
  \bibfield  {author} {\bibinfo {author} {\bibfnamefont {I.}~\bibnamefont
  {Stetcu}}, \bibinfo {author} {\bibfnamefont {S.}~\bibnamefont {Quaglioni}},
  \bibinfo {author} {\bibfnamefont {J.~L.}\ \bibnamefont {Friar}}, \bibinfo
  {author} {\bibfnamefont {A.~C.}\ \bibnamefont {Hayes}},\ and\ \bibinfo
  {author} {\bibfnamefont {P.}~\bibnamefont {Navr{\'a}til}},\ }\href@noop {}
  {\bibfield  {journal} {\bibinfo  {journal} {Physical Review C}\ }\textbf
  {\bibinfo {volume} {79}},\ \bibinfo {pages} {064001} (\bibinfo {year}
  {2009})}\BibitemShut {NoStop}%
\bibitem [{\citenamefont {Brown}\ \emph {et~al.}(1988)\citenamefont {Brown},
  \citenamefont {Etchegoyen}, \citenamefont {Rae},\ and\ \citenamefont
  {Godwin}}]{brown1988computer}%
  \BibitemOpen
  \bibfield  {author} {\bibinfo {author} {\bibfnamefont {B.}~\bibnamefont
  {Brown}}, \bibinfo {author} {\bibfnamefont {A.}~\bibnamefont {Etchegoyen}},
  \bibinfo {author} {\bibfnamefont {W.}~\bibnamefont {Rae}},\ and\ \bibinfo
  {author} {\bibfnamefont {N.}~\bibnamefont {Godwin}},\ }\href@noop {}
  {\bibfield  {journal} {\bibinfo  {journal} {MSU-NSCL Report}\ }\textbf
  {\bibinfo {volume} {524}} (\bibinfo {year} {1988})}\BibitemShut {NoStop}%
\bibitem [{\citenamefont {Warburton}\ and\ \citenamefont
  {Brown}(1992)}]{warburton1992effective}%
  \BibitemOpen
  \bibfield  {author} {\bibinfo {author} {\bibfnamefont {E.}~\bibnamefont
  {Warburton}}\ and\ \bibinfo {author} {\bibfnamefont {B.~A.}\ \bibnamefont
  {Brown}},\ }\href@noop {} {\bibfield  {journal} {\bibinfo  {journal}
  {Physical Review C}\ }\textbf {\bibinfo {volume} {46}},\ \bibinfo {pages}
  {923} (\bibinfo {year} {1992})}\BibitemShut {NoStop}%
\bibitem [{\citenamefont {Lubna}\ \emph {et~al.}(2019)\citenamefont {Lubna},
  \citenamefont {Kravvaris}, \citenamefont {Tabor}, \citenamefont {Tripathi},
  \citenamefont {Volya}, \citenamefont {Rubino}, \citenamefont {Allmond},
  \citenamefont {Abromeit}, \citenamefont {Baby},\ and\ \citenamefont
  {Hensley}}]{lubna2019structure}%
  \BibitemOpen
  \bibfield  {author} {\bibinfo {author} {\bibfnamefont {R.~S.}\ \bibnamefont
  {Lubna}}, \bibinfo {author} {\bibfnamefont {K.}~\bibnamefont {Kravvaris}},
  \bibinfo {author} {\bibfnamefont {S.~L.}\ \bibnamefont {Tabor}}, \bibinfo
  {author} {\bibfnamefont {V.}~\bibnamefont {Tripathi}}, \bibinfo {author}
  {\bibfnamefont {A.}~\bibnamefont {Volya}}, \bibinfo {author} {\bibfnamefont
  {E.}~\bibnamefont {Rubino}}, \bibinfo {author} {\bibfnamefont
  {J.}~\bibnamefont {Allmond}}, \bibinfo {author} {\bibfnamefont
  {B.}~\bibnamefont {Abromeit}}, \bibinfo {author} {\bibfnamefont
  {L.}~\bibnamefont {Baby}},\ and\ \bibinfo {author} {\bibfnamefont
  {T.}~\bibnamefont {Hensley}},\ }\href@noop {} {\bibfield  {journal} {\bibinfo
   {journal} {Physical Review C}\ }\textbf {\bibinfo {volume} {100}},\ \bibinfo
  {pages} {034308} (\bibinfo {year} {2019})}\BibitemShut {NoStop}%
\bibitem [{\citenamefont {Lubna}\ \emph {et~al.}(2020)\citenamefont {Lubna},
  \citenamefont {Kravvaris}, \citenamefont {Tabor}, \citenamefont {Tripathi},
  \citenamefont {Rubino},\ and\ \citenamefont {Volya}}]{lubna2020evolution}%
  \BibitemOpen
  \bibfield  {author} {\bibinfo {author} {\bibfnamefont {R.~S.}\ \bibnamefont
  {Lubna}}, \bibinfo {author} {\bibfnamefont {K.}~\bibnamefont {Kravvaris}},
  \bibinfo {author} {\bibfnamefont {S.~L.}\ \bibnamefont {Tabor}}, \bibinfo
  {author} {\bibfnamefont {V.}~\bibnamefont {Tripathi}}, \bibinfo {author}
  {\bibfnamefont {E.}~\bibnamefont {Rubino}},\ and\ \bibinfo {author}
  {\bibfnamefont {A.}~\bibnamefont {Volya}},\ }\href@noop {} {\bibfield
  {journal} {\bibinfo  {journal} {Physical Review Research}\ }\textbf {\bibinfo
  {volume} {2}},\ \bibinfo {pages} {043342} (\bibinfo {year}
  {2020})}\BibitemShut {NoStop}%
\bibitem [{\citenamefont {Brown}(2022)}]{brown2022nuclear}%
  \BibitemOpen
  \bibfield  {author} {\bibinfo {author} {\bibfnamefont {B.~A.}\ \bibnamefont
  {Brown}},\ }\href@noop {} {\bibfield  {journal} {\bibinfo  {journal}
  {Physics}\ }\textbf {\bibinfo {volume} {4}},\ \bibinfo {pages} {525}
  (\bibinfo {year} {2022})}\BibitemShut {NoStop}%
\bibitem [{\citenamefont {Brown}\ and\ \citenamefont
  {Richter}(2006)}]{brown2006new}%
  \BibitemOpen
  \bibfield  {author} {\bibinfo {author} {\bibfnamefont {B.~A.}\ \bibnamefont
  {Brown}}\ and\ \bibinfo {author} {\bibfnamefont {W.}~\bibnamefont
  {Richter}},\ }\href@noop {} {\bibfield  {journal} {\bibinfo  {journal}
  {Physical Review C}\ }\textbf {\bibinfo {volume} {74}},\ \bibinfo {pages}
  {034315} (\bibinfo {year} {2006})}\BibitemShut {NoStop}%
\bibitem [{\citenamefont {Levinger}\ and\ \citenamefont
  {Bethe}(1950)}]{levinger1950dipole}%
  \BibitemOpen
  \bibfield  {author} {\bibinfo {author} {\bibfnamefont {J.~S.}\ \bibnamefont
  {Levinger}}\ and\ \bibinfo {author} {\bibfnamefont {H.~A.}\ \bibnamefont
  {Bethe}},\ }\href@noop {} {\bibfield  {journal} {\bibinfo  {journal}
  {Physical Review}\ }\textbf {\bibinfo {volume} {78}},\ \bibinfo {pages} {115}
  (\bibinfo {year} {1950})}\BibitemShut {NoStop}%
\bibitem [{\citenamefont {Warburton}\ and\ \citenamefont
  {Weneser}(1969)}]{warburton1969role}%
  \BibitemOpen
  \bibfield  {author} {\bibinfo {author} {\bibfnamefont {E.~K.}\ \bibnamefont
  {Warburton}}\ and\ \bibinfo {author} {\bibfnamefont {J.}~\bibnamefont
  {Weneser}},\ }\href@noop {} {\bibfield  {journal} {\bibinfo  {journal}
  {Isospin in Nuclear Physics, ed D. H. Wilkinson, North-Holland, Amsterdam}\
  }\textbf {\bibinfo {volume} {4}},\ \bibinfo {pages} {10} (\bibinfo {year}
  {1969})}\BibitemShut {NoStop}%
\bibitem [{\citenamefont {Bohr}\ and\ \citenamefont
  {Mottelson}(1998)}]{bohr1998nuclear}%
  \BibitemOpen
  \bibfield  {author} {\bibinfo {author} {\bibfnamefont {A.~N.}\ \bibnamefont
  {Bohr}}\ and\ \bibinfo {author} {\bibfnamefont {B.~R.}\ \bibnamefont
  {Mottelson}},\ }\href@noop {} {\emph {\bibinfo {title} {Nuclear Structure (in
  2 volumes)}}}\ (\bibinfo  {publisher} {World Scientific Publishing Company},\
  \bibinfo {year} {1998})\BibitemShut {NoStop}%
\bibitem [{\citenamefont {Morinaga}(1955)}]{morinaga1955effects}%
  \BibitemOpen
  \bibfield  {author} {\bibinfo {author} {\bibfnamefont {H.}~\bibnamefont
  {Morinaga}},\ }\href@noop {} {\bibfield  {journal} {\bibinfo  {journal}
  {Physical Review}\ }\textbf {\bibinfo {volume} {97}},\ \bibinfo {pages} {444}
  (\bibinfo {year} {1955})}\BibitemShut {NoStop}%
\bibitem [{\citenamefont {Barker}\ and\ \citenamefont
  {Mann}(1957)}]{barker1957effect}%
  \BibitemOpen
  \bibfield  {author} {\bibinfo {author} {\bibfnamefont {F.}~\bibnamefont
  {Barker}}\ and\ \bibinfo {author} {\bibfnamefont {A.}~\bibnamefont {Mann}},\
  }\href@noop {} {\bibfield  {journal} {\bibinfo  {journal} {Philosophical
  Magazine}\ }\textbf {\bibinfo {volume} {2}},\ \bibinfo {pages} {5} (\bibinfo
  {year} {1957})}\BibitemShut {NoStop}%
\bibitem [{\citenamefont {Sieja}(2022)}]{Sieja2022}%
  \BibitemOpen
  \bibfield  {author} {\bibinfo {author} {\bibfnamefont {K.}~\bibnamefont
  {Sieja}},\ }\href@noop {} {\bibinfo {title} {Dipole excitations in nuclei:
  recent configuration interaction studies,}},\ \bibinfo {howpublished}
  {presented at Giant and soft modes of excitation in nuclear structure and
  astrophysics, ECT*} (\bibinfo {year} {2022})\BibitemShut {NoStop}%
\bibitem [{\citenamefont {Sieja}(2023)}]{sieja2023brink}%
  \BibitemOpen
  \bibfield  {author} {\bibinfo {author} {\bibfnamefont {K.}~\bibnamefont
  {Sieja}},\ }\href@noop {} {\bibfield  {journal} {\bibinfo  {journal} {The
  European Physical Journal A}\ }\textbf {\bibinfo {volume} {59}},\ \bibinfo
  {pages} {147} (\bibinfo {year} {2023})}\BibitemShut {NoStop}%
\bibitem [{\citenamefont {Gloeckner}\ and\ \citenamefont
  {Lawson}(1974)}]{gloeckner1974spurious}%
  \BibitemOpen
  \bibfield  {author} {\bibinfo {author} {\bibfnamefont {D.~H.}\ \bibnamefont
  {Gloeckner}}\ and\ \bibinfo {author} {\bibfnamefont {R.~D.}\ \bibnamefont
  {Lawson}},\ }\href@noop {} {\bibfield  {journal} {\bibinfo  {journal}
  {Physics Letters B}\ }\textbf {\bibinfo {volume} {53}},\ \bibinfo {pages}
  {313} (\bibinfo {year} {1974})}\BibitemShut {NoStop}%
\bibitem [{\citenamefont {Brown}(2005)}]{brown2005lecture}%
  \BibitemOpen
  \bibfield  {author} {\bibinfo {author} {\bibfnamefont {B.~A.}\ \bibnamefont
  {Brown}},\ }\href@noop {} {\bibfield  {journal} {\bibinfo  {journal}
  {National Super Conducting Cyclotron Laboratory}\ }\textbf {\bibinfo {volume}
  {11}} (\bibinfo {year} {2005})}\BibitemShut {NoStop}%
\bibitem [{\citenamefont {Neudatchin}\ \emph {et~al.}(1979)\citenamefont
  {Neudatchin}, \citenamefont {Smirnov},\ and\ \citenamefont
  {Golovanova}}]{neudatchin1979clustering}%
  \BibitemOpen
  \bibfield  {author} {\bibinfo {author} {\bibfnamefont {V.}~\bibnamefont
  {Neudatchin}}, \bibinfo {author} {\bibfnamefont {Y.~F.}\ \bibnamefont
  {Smirnov}},\ and\ \bibinfo {author} {\bibfnamefont {N.}~\bibnamefont
  {Golovanova}},\ }\href@noop {} {\bibfield  {journal} {\bibinfo  {journal}
  {Adv. Nucl. Phys.;(United States)}\ }\textbf {\bibinfo {volume} {11}}
  (\bibinfo {year} {1979})}\BibitemShut {NoStop}%
\bibitem [{\citenamefont {He}\ \emph {et~al.}(2014)\citenamefont {He},
  \citenamefont {Ma}, \citenamefont {Cao}, \citenamefont {Cai}, \citenamefont
  {Zhang} \emph {et~al.}}]{he2014giant}%
  \BibitemOpen
  \bibfield  {author} {\bibinfo {author} {\bibfnamefont {W.~B.}\ \bibnamefont
  {He}}, \bibinfo {author} {\bibfnamefont {Y.~G.}\ \bibnamefont {Ma}}, \bibinfo
  {author} {\bibfnamefont {X.~G.}\ \bibnamefont {Cao}}, \bibinfo {author}
  {\bibfnamefont {X.~Z.}\ \bibnamefont {Cai}}, \bibinfo {author} {\bibfnamefont
  {G.~Q.}\ \bibnamefont {Zhang}}, \emph {et~al.},\ }\href@noop {} {\bibfield
  {journal} {\bibinfo  {journal} {Physical Review Letters}\ }\textbf {\bibinfo
  {volume} {113}},\ \bibinfo {pages} {032506} (\bibinfo {year}
  {2014})}\BibitemShut {NoStop}%
\bibitem [{\citenamefont {Smilansky}\ \emph {et~al.}(1972)\citenamefont
  {Smilansky}, \citenamefont {Povh},\ and\ \citenamefont
  {Traxel}}]{smilansky1972role}%
  \BibitemOpen
  \bibfield  {author} {\bibinfo {author} {\bibfnamefont {U.}~\bibnamefont
  {Smilansky}}, \bibinfo {author} {\bibfnamefont {B.}~\bibnamefont {Povh}},\
  and\ \bibinfo {author} {\bibfnamefont {K.}~\bibnamefont {Traxel}},\
  }\href@noop {} {\bibfield  {journal} {\bibinfo  {journal} {Physics Letters
  B}\ }\textbf {\bibinfo {volume} {38}},\ \bibinfo {pages} {293} (\bibinfo
  {year} {1972})}\BibitemShut {NoStop}%
\bibitem [{\citenamefont {Weller}\ \emph {et~al.}(1985)\citenamefont {Weller},
  \citenamefont {Egelhof}, \citenamefont {{\v{C}}aplar}, \citenamefont
  {Karban}, \citenamefont {Kr{\"a}mer}, \citenamefont {M{\"o}bius},
  \citenamefont {Moroz}, \citenamefont {Rusek}, \citenamefont {Steffens},
  \citenamefont {Tungate} \emph {et~al.}}]{weller1985electromagnetic}%
  \BibitemOpen
  \bibfield  {author} {\bibinfo {author} {\bibfnamefont {A.}~\bibnamefont
  {Weller}}, \bibinfo {author} {\bibfnamefont {P.}~\bibnamefont {Egelhof}},
  \bibinfo {author} {\bibfnamefont {R.}~\bibnamefont {{\v{C}}aplar}}, \bibinfo
  {author} {\bibfnamefont {O.}~\bibnamefont {Karban}}, \bibinfo {author}
  {\bibfnamefont {D.}~\bibnamefont {Kr{\"a}mer}}, \bibinfo {author}
  {\bibfnamefont {K.-H.}\ \bibnamefont {M{\"o}bius}}, \bibinfo {author}
  {\bibfnamefont {Z.}~\bibnamefont {Moroz}}, \bibinfo {author} {\bibfnamefont
  {K.}~\bibnamefont {Rusek}}, \bibinfo {author} {\bibfnamefont
  {E.}~\bibnamefont {Steffens}}, \bibinfo {author} {\bibfnamefont
  {G.}~\bibnamefont {Tungate}}, \emph {et~al.},\ }\href@noop {} {\bibfield
  {journal} {\bibinfo  {journal} {Physical Review Letters}\ }\textbf {\bibinfo
  {volume} {55}},\ \bibinfo {pages} {480} (\bibinfo {year} {1985})}\BibitemShut
  {NoStop}%
\bibitem [{\citenamefont {Barker}\ and\ \citenamefont
  {Woods}(1989)}]{barke1989investigation}%
  \BibitemOpen
  \bibfield  {author} {\bibinfo {author} {\bibfnamefont {F.~C.}\ \bibnamefont
  {Barker}}\ and\ \bibinfo {author} {\bibfnamefont {C.~L.}\ \bibnamefont
  {Woods}},\ }\href@noop {} {\bibfield  {journal} {\bibinfo  {journal}
  {Australian Journal of Physics}\ }\textbf {\bibinfo {volume} {42}},\ \bibinfo
  {pages} {233} (\bibinfo {year} {1989})}\BibitemShut {NoStop}%
\bibitem [{\citenamefont {Barker}(1984)}]{barker1984decay}%
  \BibitemOpen
  \bibfield  {author} {\bibinfo {author} {\bibfnamefont {F.~C.}\ \bibnamefont
  {Barker}},\ }\href@noop {} {\bibfield  {journal} {\bibinfo  {journal}
  {Australian Journal of Physics}\ }\textbf {\bibinfo {volume} {37}},\ \bibinfo
  {pages} {267} (\bibinfo {year} {1984})}\BibitemShut {NoStop}%
\bibitem [{\citenamefont {Ikeda}\ \emph {et~al.}(1968)\citenamefont {Ikeda},
  \citenamefont {Takigawa},\ and\ \citenamefont
  {Horiuchi}}]{ikeda1968systematic}%
  \BibitemOpen
  \bibfield  {author} {\bibinfo {author} {\bibfnamefont {K.}~\bibnamefont
  {Ikeda}}, \bibinfo {author} {\bibfnamefont {N.}~\bibnamefont {Takigawa}},\
  and\ \bibinfo {author} {\bibfnamefont {H.}~\bibnamefont {Horiuchi}},\
  }\href@noop {} {\bibfield  {journal} {\bibinfo  {journal} {Progress of
  Theoretical Physics Supplement}\ }\textbf {\bibinfo {volume} {68}},\ \bibinfo
  {pages} {464} (\bibinfo {year} {1968})}\BibitemShut {NoStop}%
\bibitem [{\citenamefont {Kanada-En'yo}\ and\ \citenamefont
  {Horiuchi}(1995)}]{kanada1995clustering}%
  \BibitemOpen
  \bibfield  {author} {\bibinfo {author} {\bibfnamefont {Y.}~\bibnamefont
  {Kanada-En'yo}}\ and\ \bibinfo {author} {\bibfnamefont {H.}~\bibnamefont
  {Horiuchi}},\ }\href@noop {} {\bibfield  {journal} {\bibinfo  {journal}
  {Progress of Theoretical Physics}\ }\textbf {\bibinfo {volume} {93}},\
  \bibinfo {pages} {115} (\bibinfo {year} {1995})}\BibitemShut {NoStop}%
\bibitem [{\citenamefont {von Oertzen}\ \emph {et~al.}(2006)\citenamefont {von
  Oertzen}, \citenamefont {Freer},\ and\ \citenamefont
  {Kanada-En’yo}}]{von2006nuclear}%
  \BibitemOpen
  \bibfield  {author} {\bibinfo {author} {\bibfnamefont {W.}~\bibnamefont {von
  Oertzen}}, \bibinfo {author} {\bibfnamefont {M.}~\bibnamefont {Freer}},\ and\
  \bibinfo {author} {\bibfnamefont {Y.}~\bibnamefont {Kanada-En’yo}},\
  }\href@noop {} {\bibfield  {journal} {\bibinfo  {journal} {Physics Reports}\
  }\textbf {\bibinfo {volume} {432}},\ \bibinfo {pages} {43} (\bibinfo {year}
  {2006})}\BibitemShut {NoStop}%
\bibitem [{\citenamefont {Freer}(2007)}]{freer2007clustered}%
  \BibitemOpen
  \bibfield  {author} {\bibinfo {author} {\bibfnamefont {M.}~\bibnamefont
  {Freer}},\ }\href@noop {} {\bibfield  {journal} {\bibinfo  {journal} {Reports
  on Progress in Physics}\ }\textbf {\bibinfo {volume} {70}},\ \bibinfo {pages}
  {2149} (\bibinfo {year} {2007})}\BibitemShut {NoStop}%
\bibitem [{\citenamefont {Ebran}\ \emph {et~al.}(2012)\citenamefont {Ebran},
  \citenamefont {Khan}, \citenamefont {Nik{\v{s}}i{\'c}},\ and\ \citenamefont
  {Vretenar}}]{ebran2012atomic}%
  \BibitemOpen
  \bibfield  {author} {\bibinfo {author} {\bibfnamefont {J.-P.}\ \bibnamefont
  {Ebran}}, \bibinfo {author} {\bibfnamefont {E.}~\bibnamefont {Khan}},
  \bibinfo {author} {\bibfnamefont {T.}~\bibnamefont {Nik{\v{s}}i{\'c}}},\ and\
  \bibinfo {author} {\bibfnamefont {D.}~\bibnamefont {Vretenar}},\ }\href@noop
  {} {\bibfield  {journal} {\bibinfo  {journal} {Nature}\ }\textbf {\bibinfo
  {volume} {487}},\ \bibinfo {pages} {341} (\bibinfo {year}
  {2012})}\BibitemShut {NoStop}%
\bibitem [{\citenamefont {Zhou}\ \emph {et~al.}(2012)\citenamefont {Zhou},
  \citenamefont {Ren}, \citenamefont {Xu}, \citenamefont {Funaki},
  \citenamefont {Yamada}, \citenamefont {Tohsaki}, \citenamefont {Horiuchi},
  \citenamefont {Schuck},\ and\ \citenamefont {R{\"o}pke}}]{zhou2012new}%
  \BibitemOpen
  \bibfield  {author} {\bibinfo {author} {\bibfnamefont {B.}~\bibnamefont
  {Zhou}}, \bibinfo {author} {\bibfnamefont {Z.}~\bibnamefont {Ren}}, \bibinfo
  {author} {\bibfnamefont {C.}~\bibnamefont {Xu}}, \bibinfo {author}
  {\bibfnamefont {Y.}~\bibnamefont {Funaki}}, \bibinfo {author} {\bibfnamefont
  {T.}~\bibnamefont {Yamada}}, \bibinfo {author} {\bibfnamefont
  {A.}~\bibnamefont {Tohsaki}}, \bibinfo {author} {\bibfnamefont
  {H.}~\bibnamefont {Horiuchi}}, \bibinfo {author} {\bibfnamefont
  {P.}~\bibnamefont {Schuck}},\ and\ \bibinfo {author} {\bibfnamefont
  {G.}~\bibnamefont {R{\"o}pke}},\ }\href@noop {} {\bibfield  {journal}
  {\bibinfo  {journal} {Physical Review C}\ }\textbf {\bibinfo {volume} {86}},\
  \bibinfo {pages} {014301} (\bibinfo {year} {2012})}\BibitemShut {NoStop}%
\bibitem [{\citenamefont {Kuehner}\ \emph {et~al.}(1982)\citenamefont
  {Kuehner}, \citenamefont {Spear}, \citenamefont {Vermeer}, \citenamefont
  {Esat}, \citenamefont {Baxter},\ and\ \citenamefont
  {Hinds}}]{kuehner1982measurement}%
  \BibitemOpen
  \bibfield  {author} {\bibinfo {author} {\bibfnamefont {J.~A.}\ \bibnamefont
  {Kuehner}}, \bibinfo {author} {\bibfnamefont {R.~H.}\ \bibnamefont {Spear}},
  \bibinfo {author} {\bibfnamefont {W.~J.}\ \bibnamefont {Vermeer}}, \bibinfo
  {author} {\bibfnamefont {M.~T.}\ \bibnamefont {Esat}}, \bibinfo {author}
  {\bibfnamefont {A.~M.}\ \bibnamefont {Baxter}},\ and\ \bibinfo {author}
  {\bibfnamefont {S.}~\bibnamefont {Hinds}},\ }\href@noop {} {\bibfield
  {journal} {\bibinfo  {journal} {Physics Letters B}\ }\textbf {\bibinfo
  {volume} {115}},\ \bibinfo {pages} {437} (\bibinfo {year}
  {1982})}\BibitemShut {NoStop}%
\bibitem [{\citenamefont {Tian}\ \emph {et~al.}(2014)\citenamefont {Tian},
  \citenamefont {Cui}, \citenamefont {Zheng},\ and\ \citenamefont
  {Wang}}]{tian2014effect}%
  \BibitemOpen
  \bibfield  {author} {\bibinfo {author} {\bibfnamefont {J.}~\bibnamefont
  {Tian}}, \bibinfo {author} {\bibfnamefont {H.}~\bibnamefont {Cui}}, \bibinfo
  {author} {\bibfnamefont {K.}~\bibnamefont {Zheng}},\ and\ \bibinfo {author}
  {\bibfnamefont {N.}~\bibnamefont {Wang}},\ }\href@noop {} {\bibfield
  {journal} {\bibinfo  {journal} {Physical Review C}\ }\textbf {\bibinfo
  {volume} {90}},\ \bibinfo {pages} {024313} (\bibinfo {year}
  {2014})}\BibitemShut {NoStop}%
\bibitem [{\citenamefont {Myers}\ and\ \citenamefont
  {Swiatecki}(1969)}]{myers1969average}%
  \BibitemOpen
  \bibfield  {author} {\bibinfo {author} {\bibfnamefont {W.~D.}\ \bibnamefont
  {Myers}}\ and\ \bibinfo {author} {\bibfnamefont {W.~J.}\ \bibnamefont
  {Swiatecki}},\ }\href@noop {} {\bibfield  {journal} {\bibinfo  {journal}
  {Annals of Physics}\ }\textbf {\bibinfo {volume} {55}},\ \bibinfo {pages}
  {395} (\bibinfo {year} {1969})}\BibitemShut {NoStop}%
\bibitem [{\citenamefont {M{\"o}ller}\ \emph {et~al.}(2016)\citenamefont
  {M{\"o}ller}, \citenamefont {Sierk}, \citenamefont {Ichikawa},\ and\
  \citenamefont {Sagawa}}]{moller2016nuclear}%
  \BibitemOpen
  \bibfield  {author} {\bibinfo {author} {\bibfnamefont {P.}~\bibnamefont
  {M{\"o}ller}}, \bibinfo {author} {\bibfnamefont {A.~J.}\ \bibnamefont
  {Sierk}}, \bibinfo {author} {\bibfnamefont {T.}~\bibnamefont {Ichikawa}},\
  and\ \bibinfo {author} {\bibfnamefont {H.}~\bibnamefont {Sagawa}},\
  }\href@noop {} {\bibfield  {journal} {\bibinfo  {journal} {Atomic Data and
  Nuclear Data Tables}\ }\textbf {\bibinfo {volume} {109}},\ \bibinfo {pages}
  {1} (\bibinfo {year} {2016})}\BibitemShut {NoStop}%
\bibitem [{\citenamefont {Thomas}(1925)}]{thomas1925zahl}%
  \BibitemOpen
  \bibfield  {author} {\bibinfo {author} {\bibfnamefont {W.}~\bibnamefont
  {Thomas}},\ }\href@noop {} {\bibfield  {journal} {\bibinfo  {journal}
  {Naturwissenschaften}\ }\textbf {\bibinfo {volume} {13}},\ \bibinfo {pages}
  {627} (\bibinfo {year} {1925})}\BibitemShut {NoStop}%
\bibitem [{\citenamefont {Ladenburg}\ and\ \citenamefont
  {Reiche}(1923)}]{ladenburg1923absorption}%
  \BibitemOpen
  \bibfield  {author} {\bibinfo {author} {\bibfnamefont {R.}~\bibnamefont
  {Ladenburg}}\ and\ \bibinfo {author} {\bibfnamefont {F.}~\bibnamefont
  {Reiche}},\ }\href@noop {} {\bibfield  {journal} {\bibinfo  {journal}
  {Naturwissenschaften}\ }\textbf {\bibinfo {volume} {11}},\ \bibinfo {pages}
  {584} (\bibinfo {year} {1923})}\BibitemShut {NoStop}%
\bibitem [{\citenamefont {Reiche}\ and\ \citenamefont
  {Thomas}(1925)}]{reiche1925zahl}%
  \BibitemOpen
  \bibfield  {author} {\bibinfo {author} {\bibfnamefont {F.}~\bibnamefont
  {Reiche}}\ and\ \bibinfo {author} {\bibfnamefont {W.}~\bibnamefont
  {Thomas}},\ }\href@noop {} {\bibfield  {journal} {\bibinfo  {journal}
  {Zeitschrift f{\"u}r Physik}\ }\textbf {\bibinfo {volume} {34}},\ \bibinfo
  {pages} {510} (\bibinfo {year} {1925})}\BibitemShut {NoStop}%
\bibitem [{\citenamefont {Kuhn}(1925)}]{kuhn1925gesamtstarke}%
  \BibitemOpen
  \bibfield  {author} {\bibinfo {author} {\bibfnamefont {W.}~\bibnamefont
  {Kuhn}},\ }\href@noop {} {\bibfield  {journal} {\bibinfo  {journal}
  {Zeitschrift f{\"u}r Physik}\ }\textbf {\bibinfo {volume} {33}},\ \bibinfo
  {pages} {408} (\bibinfo {year} {1925})}\BibitemShut {NoStop}%
\bibitem [{\citenamefont {Bertsch}(1972)}]{bertsch1972practitioner}%
  \BibitemOpen
  \bibfield  {author} {\bibinfo {author} {\bibfnamefont {G.~F.}\ \bibnamefont
  {Bertsch}},\ }\href@noop {} {\emph {\bibinfo {title} {The practitioner's
  shell model}}}\ (\bibinfo  {publisher} {North Holland Publishing Company},\
  \bibinfo {year} {1972})\BibitemShut {NoStop}%
\bibitem [{\citenamefont {Blomqvist}\ and\ \citenamefont
  {Molinari}(1968)}]{blomqvist1968collective}%
  \BibitemOpen
  \bibfield  {author} {\bibinfo {author} {\bibfnamefont {J.}~\bibnamefont
  {Blomqvist}}\ and\ \bibinfo {author} {\bibfnamefont {A.}~\bibnamefont
  {Molinari}},\ }\href@noop {} {\bibfield  {journal} {\bibinfo  {journal}
  {Nuclear Physics A}\ }\textbf {\bibinfo {volume} {106}},\ \bibinfo {pages}
  {545} (\bibinfo {year} {1968})}\BibitemShut {NoStop}%
\bibitem [{\citenamefont {Fearick}\ \emph {et~al.}(2018)\citenamefont
  {Fearick}, \citenamefont {Erler}, \citenamefont {Matsubara}, \citenamefont
  {von Neumann-Cosel}, \citenamefont {Richter}, \citenamefont {Roth},\ and\
  \citenamefont {Tamii}}]{fearick2018origin}%
  \BibitemOpen
  \bibfield  {author} {\bibinfo {author} {\bibfnamefont {R.}~\bibnamefont
  {Fearick}}, \bibinfo {author} {\bibfnamefont {B.}~\bibnamefont {Erler}},
  \bibinfo {author} {\bibfnamefont {H.}~\bibnamefont {Matsubara}}, \bibinfo
  {author} {\bibfnamefont {P.}~\bibnamefont {von Neumann-Cosel}}, \bibinfo
  {author} {\bibfnamefont {A.}~\bibnamefont {Richter}}, \bibinfo {author}
  {\bibfnamefont {R.}~\bibnamefont {Roth}},\ and\ \bibinfo {author}
  {\bibfnamefont {A.}~\bibnamefont {Tamii}},\ }\href@noop {} {\bibfield
  {journal} {\bibinfo  {journal} {Physical Review C}\ }\textbf {\bibinfo
  {volume} {97}},\ \bibinfo {pages} {044325} (\bibinfo {year}
  {2018})}\BibitemShut {NoStop}%
\bibitem [{\citenamefont {Tamii}\ \emph {et~al.}(2022)\citenamefont {Tamii},
  \citenamefont {Pellegri}, \citenamefont {S{\"o}derstr{\"o}m}, \citenamefont
  {Allard}, \citenamefont {Goriely}, \citenamefont {Inakura}, \citenamefont
  {Khan}, \citenamefont {Kido}, \citenamefont {Kimura}, \citenamefont
  {Litvinova} \emph {et~al.}}]{tamii2022pandora}%
  \BibitemOpen
  \bibfield  {author} {\bibinfo {author} {\bibfnamefont {A.}~\bibnamefont
  {Tamii}}, \bibinfo {author} {\bibfnamefont {L.}~\bibnamefont {Pellegri}},
  \bibinfo {author} {\bibfnamefont {P.-A.}\ \bibnamefont {S{\"o}derstr{\"o}m}},
  \bibinfo {author} {\bibfnamefont {D.}~\bibnamefont {Allard}}, \bibinfo
  {author} {\bibfnamefont {S.}~\bibnamefont {Goriely}}, \bibinfo {author}
  {\bibfnamefont {T.}~\bibnamefont {Inakura}}, \bibinfo {author} {\bibfnamefont
  {E.}~\bibnamefont {Khan}}, \bibinfo {author} {\bibfnamefont {E.}~\bibnamefont
  {Kido}}, \bibinfo {author} {\bibfnamefont {M.}~\bibnamefont {Kimura}},
  \bibinfo {author} {\bibfnamefont {E.}~\bibnamefont {Litvinova}}, \emph
  {et~al.},\ }\href@noop {} {\bibfield  {journal} {\bibinfo  {journal} {arXiv
  preprint arXiv:2211.03986}\ } (\bibinfo {year} {2022})}\BibitemShut {NoStop}%
\bibitem [{\citenamefont {Ferentz}\ \emph {et~al.}(1953)\citenamefont
  {Ferentz}, \citenamefont {Gell-Mann},\ and\ \citenamefont
  {Pines}}]{ferentz1953giant}%
  \BibitemOpen
  \bibfield  {author} {\bibinfo {author} {\bibfnamefont {M.}~\bibnamefont
  {Ferentz}}, \bibinfo {author} {\bibfnamefont {M.}~\bibnamefont {Gell-Mann}},\
  and\ \bibinfo {author} {\bibfnamefont {D.}~\bibnamefont {Pines}},\
  }\href@noop {} {\bibfield  {journal} {\bibinfo  {journal} {Physical Review}\
  }\textbf {\bibinfo {volume} {92}},\ \bibinfo {pages} {836} (\bibinfo {year}
  {1953})}\BibitemShut {NoStop}%
\bibitem [{\citenamefont {Johnson}\ and\ \citenamefont
  {Teller}(1955)}]{johnson1955classical}%
  \BibitemOpen
  \bibfield  {author} {\bibinfo {author} {\bibfnamefont {M.~H.}\ \bibnamefont
  {Johnson}}\ and\ \bibinfo {author} {\bibfnamefont {E.}~\bibnamefont
  {Teller}},\ }\href@noop {} {\bibfield  {journal} {\bibinfo  {journal}
  {Physical Review}\ }\textbf {\bibinfo {volume} {98}},\ \bibinfo {pages} {783}
  (\bibinfo {year} {1955})}\BibitemShut {NoStop}%
\bibitem [{\citenamefont {Weisskopf}(1957)}]{weisskopf1957problem}%
  \BibitemOpen
  \bibfield  {author} {\bibinfo {author} {\bibfnamefont {V.~F.}\ \bibnamefont
  {Weisskopf}},\ }\href@noop {} {\bibfield  {journal} {\bibinfo  {journal}
  {Nuclear Physics}\ }\textbf {\bibinfo {volume} {3}},\ \bibinfo {pages} {423}
  (\bibinfo {year} {1957})}\BibitemShut {NoStop}%
\bibitem [{\citenamefont {Rand}(1957)}]{rand1957appreciation}%
  \BibitemOpen
  \bibfield  {author} {\bibinfo {author} {\bibfnamefont {S.}~\bibnamefont
  {Rand}},\ }\href@noop {} {\bibfield  {journal} {\bibinfo  {journal} {Physical
  Review}\ }\textbf {\bibinfo {volume} {107}},\ \bibinfo {pages} {208}
  (\bibinfo {year} {1957})}\BibitemShut {NoStop}%
\bibitem [{\citenamefont {Brenig}(1965)}]{brenig1965advances}%
  \BibitemOpen
  \bibfield  {author} {\bibinfo {author} {\bibfnamefont {W.}~\bibnamefont
  {Brenig}},\ }\href@noop {} {\bibfield  {journal} {\bibinfo  {journal}
  {Academic Press, New York, New York}\ }\textbf {\bibinfo {volume} {1}},\
  \bibinfo {pages} {59} (\bibinfo {year} {1965})}\BibitemShut {NoStop}%
\bibitem [{\citenamefont {Sarma}\ and\ \citenamefont
  {Srivastava}(2022)}]{sarma2022ab}%
  \BibitemOpen
  \bibfield  {author} {\bibinfo {author} {\bibfnamefont {C.}~\bibnamefont
  {Sarma}}\ and\ \bibinfo {author} {\bibfnamefont {P.~C.}\ \bibnamefont
  {Srivastava}},\ }\href@noop {} {\bibfield  {journal} {\bibinfo  {journal}
  {arXiv preprint arXiv:2208.00816}\ } (\bibinfo {year} {2022})}\BibitemShut
  {NoStop}%
\bibitem [{\citenamefont {Eichler}(1964)}]{eichler1964second}%
  \BibitemOpen
  \bibfield  {author} {\bibinfo {author} {\bibfnamefont {J.}~\bibnamefont
  {Eichler}},\ }\href@noop {} {\bibfield  {journal} {\bibinfo  {journal}
  {Physical Review}\ }\textbf {\bibinfo {volume} {133}},\ \bibinfo {pages}
  {B1162} (\bibinfo {year} {1964})}\BibitemShut {NoStop}%
\end{thebibliography}%

\end{document}